\documentclass{emulateapj}
\usepackage{times,mathptmx}

\usepackage{graphicx}

\newcommand{\hbeta}{\mbox{H$\beta$}}
\newcommand{\halpha}{\mbox{H$\alpha$}}
\newcommand{\hi}{\hbox{H {\sc i}}}
\newcommand{\hii}{\hbox{H {\sc ii}}}
\newcommand{\mgi}{\hbox{Mg {\sc i}}}
\newcommand{\mgii}{\hbox{Mg {\sc ii}}}
\newcommand{\oii}{\hbox{[O {\sc ii}]}}
\newcommand{\oiii}{\hbox{[O {\sc iii}]}}
\newcommand{\neiii}{\hbox{[Ne {\sc iii}]}}

\newcommand{\kms}{\hbox{$\rm {km}~\rm s^{-1}$}}

\newcommand{\um}{\hbox{$\mu$m}}
\newcommand{\etal}{et al.\,}
\newcommand{\lir}{\hbox{$L_{IR}$}}
\newcommand{\lsun}{\hbox{$L_\odot$}}
\newcommand{\msun}{\hbox{$M_\odot$}}
\newcommand{\msunyr}{\hbox{$M_\odot~\rm{yr}^{-1}$}}
\newcommand{\msunyrkpcsq}{\hbox{$M_\odot~\rm{yr}^{-1}\ \rm{kpc}^{-2}$}}
\newcommand{\mujy}{\hbox{$\mu$Jy}}

\newcommand{\cmsq}{\hbox{cm$^{-2}$}}

\shortauthors{Weiner et al.}
\shorttitle{Outflows from Star-Forming Galaxies}

\slugcomment{ApJ in press}

\begin{document}

\title{Ubiquitous Outflows in DEEP2 Spectra of Star-Forming Galaxies at z=1.4}

\author{Benjamin J. Weiner\altaffilmark{1,2},
Alison L. Coil\altaffilmark{1,3,4},
Jason X. Prochaska\altaffilmark{5}, 
Jeffrey A. Newman\altaffilmark{6}, 
Michael C. Cooper\altaffilmark{1,7}, 
Kevin Bundy\altaffilmark{8}, 
Christopher J. Conselice\altaffilmark{9}, 
Aaron A. Dutton\altaffilmark{5}, 
S. M. Faber\altaffilmark{5}, 
David C. Koo\altaffilmark{5}, 
Jennifer M. Lotz\altaffilmark{10,11},
G.H. Rieke\altaffilmark{1}, 
K.H.R. Rubin\altaffilmark{5}
}

\altaffiltext{1}{Steward Observatory, 933 N. Cherry St., University of Arizona, Tucson, AZ 85721}
\altaffiltext{2}{{\tt bjw@as.arizona.edu}}
\altaffiltext{3}{Hubble Fellow}
\altaffiltext{4}{Department of Physics, and CASS, University of California, San Diego, La Jolla, CA 92093-0424}
\altaffiltext{5}{UCO/Lick Observatory, University of California, Santa Cruz, 
Santa Cruz, CA 95064}
\altaffiltext{6}{Dept. of Physics and Astronomy, University of Pittsburgh, Pittsburgh, PA 15260}
\altaffiltext{7}{Spitzer Fellow}
\altaffiltext{8}{Department of Astronomy and Astrophysics, University of Toronto, Toronto, ON M5S 3H4, CA}
\altaffiltext{9}{School of Physics and Astronomy, University of Nottingham, Nottingham, UK}
\altaffiltext{10}{National Optical Astronomy Observatories, Tucson, AZ 85719}
\altaffiltext{11}{Leo Goldberg Fellow}

\begin{abstract}

Galactic winds are a prime suspect for the metal enrichment of the
intergalactic medium and may have a strong influence on the chemical
evolution of galaxies and the nature of QSO absorption line systems.
We use a sample of 1406 galaxy
spectra at $z\sim 1.4$ from the DEEP2 redshift survey to show
that blueshifted \mgii\ $\lambda\lambda$ 2796, 2803 \AA\ 
absorption is ubiquitous in starforming galaxies at this epoch.  
This is the first detection of frequent outflowing galactic winds at
$z\sim 1$.  The presence and depth of absorption are independent
of AGN spectral signatures or galaxy morphology; major mergers are 
not a prerequisite for driving a galactic wind from massive galaxies.  
Outflows are found in coadded spectra of galaxies spanning a range
of $30\times$ in stellar mass and $10\times$ in star formation rate (SFR),
calibrated from $K$-band and from MIPS IR fluxes.  
The outflows have column densities of order $N_H \sim 10^{20} \cmsq$
and characteristic velocities of $\sim 300-500$ km/sec, with
absorption seen out to 1000 km/sec in the most massive, highest
SFR galaxies.  The velocities suggest that the outflowing gas can
escape into the IGM and that massive galaxies can produce cosmologically 
and chemically significant outflows.  Both the \mgii\ equivalent width
and the outflow velocity are larger for galaxies of higher stellar mass
and SFR, with $V_{wind} \sim SFR^{0.3}$, similar
to the scaling in low redshift IR-luminous galaxies.  
The high frequency of outflows in the star-forming galaxy population at 
$z\sim 1$ indicates that galactic winds occur in the progenitors of 
massive spirals as well as those of ellipticals.  The
increase of outflow velocity with mass and SFR constrains
theoretical models of galaxy evolution that include feedback 
from galactic winds, and may favor momentum-driven models
for the wind physics.

\end{abstract}

\keywords{galaxies: high-redshift --- intergalactic medium --- 
galaxies: evolution --- ultraviolet: ISM}

\section{Introduction}

As galaxies form, they accrete gas from infalling objects
and from the intergalactic medium (IGM).  The process of galaxy
formation also is apparently not 100\% efficient at getting gas
into galaxies, as some
galaxies are observed to drive galactic winds or outflows,
expelling gas, possibly to escape the galaxy itself.
The baryonic mass in local disks compared to their dark matter
halo mass is lower than the global value, which may point to
past episodes of gas removal by outflows (e.g. Fukugita, Hogan \& Peebles
1998; Mo \& Mao 2004; Dutton \etal\ 2007).
Galactic winds can be driven by stellar winds, supernovae, or
active galactic nuclei; observationally, in local galaxies
winds are a general consequence of high star formation
rates or star formation surface density
(e.g. Lehnert \& Heckman 1996).  Outflows may play a 
large role in regulating the baryon and metal content of galaxies,
and in enriching the intergalactic medium.  
The frequency of outflows and their dependence on host 
galaxy properties such as mass and star formation rate (SFR) are
key ingredients for modeling the evolution of the IGM and the 
chemical evolution and gas processing of galaxies.  Observations
of winds and their implications are discussed in the 
review by Veilleux, Cecil \& Bland-Hawthorn (2005).

In nearby galaxies, strong winds driven by supernovae
are seen mainly in starbursting
dwarfs, starburst galaxies such as M82, and luminous and
ultraluminous infrared galaxies (LIRGs and ULIRGs); the latter
are frequently mergers and often host active galactic nuclei (AGN).  
Winds are often found in galaxies above a SFR surface density threshold
of $\sim 0.1 \msunyrkpcsq$ (Heckman 2002), while large-scale
outflows are not seen in local ordinary disk galaxies.  The
Milky Way has a nuclear wind that would not be detectable 
outside the Local Group (Veilleux \etal\ 2005).  It is not
clear whether a galaxy such as the Milky Way has had a
galactic-scale outflow in the past.  We do know that the global SFR
density was $\sim 10\times$ higher at $z=1$ than today
(e.g. Lilly \etal\ 1996; Hopkins 2004).  Noeske \etal\ (2007a) 
showed that the higher SFR is due to higher rates at $z=1$
in starforming galaxies across a wide mass range, rather than
an increased frequency of starbursts in a fraction of galaxies.  
It is thus possible that
a large fraction of the galaxy population drove outflows in the past;
that is one question this paper seeks to address.

The dependence of outflows on galaxy mass is also of great
interest.  Larson (1974) showed that supernova-driven winds 
could heat and eject 
interstellar gas during galaxy formation, and that this
process could be more effective in smaller galaxies.
The influential theoretical treatment of Dekel \& Silk (1986)
found that winds could remove a significant amount of
gas from low-mass galaxies, $V_{vir} \lesssim 100$ \kms,
resolving several issues in cold dark matter galaxy formation.
A number of authors have simulated the effect of winds
on the ISM, e.g. Mac Low \& Ferrara (1999).  Their simulations
suggested that winds do not entirely clear ISM gas except at 
extremely low masses, but that they are efficient at 
preferentially ejecting metals.  However, other treatments
of wind physics suggest that winds in massive galaxies in a high-SFR
phase could limit the SFR and eject a substantial amount
of gas (Murray, Quataert \& Thompson 2005).  The physical
conditions in the wind fluid and the mass-loading of cold gas
into the hot wind are quite difficult to constrain from
observations (e.g. Strickland \& Stevens 2000; Murray \etal\ 2005).
It remains open whether metals are ejected and the IGM enriched 
preferentially by low or high mass galaxies, and the degree to
which AGN-driven winds contribute.

Fossil evidence of the effect of winds comes from studies of 
the mass-metallicity relation in local samples
and in the Sloan Digital Sky Survey.  These studies show that low-mass 
galaxies have low effective yields, indicating that winds have removed
metals from them; the behavior of the 
effective yield is evidence for past winds in galaxies of stellar mass
$M_* < 10^{10} \msun$ (Garnett 2002; Tremonti \etal\ 2004).  
Dalcanton (2007) showed that star
formation after an outflow episode is efficient at returning
the measured metal yield to its nominal value.  Therefore, winds
must preferentially carry away enriched gas to reduce the 
yield in low-mass, high gas-fraction galaxies.  However, because
even enriched outflows will not substantially change the observed 
effective yield in high-mass galaxies, the observations
do not rule out winds at higher masses (Tremonti \etal\ 2004).
The behavior of the effective yield
does not provide evidence either for or against past winds 
in high mass galaxies.  

Direct evidence for galactic winds in the local universe comes
from a variety of observations including X-rays, the morphology 
and kinematics of interstellar emission lines, and outflow
kinematics in interstellar absorption lines (Veilleux
\etal\ 2005 and references therein).  Blueshifted absorption 
lines are the wind signatures
that are most straightforward to observe and interpret in distant 
galaxies, and the method we use in this paper.  

In local galaxies, Phillips (1993) found blueshifted Na I 
absorption spread across the disk of the starforming galaxy NGC 1808,
associated with extraplanar dust plumes.
Blueshifted Na I absorption was observed in 
12 of 18 infrared-luminous starburst galaxies (Heckman \etal\ 
2000), with outflow velocities of 100--600 \kms.  Further surveys
have confirmed that blueshifted Na I absorption is common in
IR-bright starburst galaxies (Rupke, Veilleux \& Sanders 2002, 2005a,b;
Martin 2005, 2006) and is also found in dwarf starbursts
(Schwartz \& Martin 2004).  Low-ionization UV lines behave 
similarly to the Na I lines (Schwartz \etal\ 2006).

Tremonti, Moustakas \& Diamond-Stanic (2007) detected blueshifted 
absorption in \mgii\ in $z\sim 0.6$ very luminous post-starburst galaxies.
The outflow velocities are very large, 500-2000 \kms, leading them to suggest
these are relics of AGN-driven outflows.
Sato \etal\ (2008) have found blueshifted Na I absorption in
red-sequence galaxies at $z \sim 0.4$ in the DEEP2 survey;
they find a higher frequency of outflows at greater infrared
luminosity, but also find some outflows in quiescent galaxies
that show signs of being post-starbursts.

At high redshift, $z \sim 3$, outflows with typical velocities 
of 250 \kms\ are found in rapidly starforming galaxies selected 
by the Lyman-break technique.  The gravitationally lensed
galaxy cB58 (Pettini \etal\ 2000, 2002) and the composite spectrum
of a large sample of Lyman-break galaxies (Shapley \etal\ 2003)
show blueshifted outflows in both low- and high-ionization UV 
interstellar lines such
as Si II and C IV.  These early, rapidly starforming galaxies
are commonly thought to be progenitors of today's massive
ellipticals (e.g. Adelberger \etal\ 2005), although this point
remains controversial (e.g. Conroy \etal\ 2008).  At $z>3$,  Lyman-$\alpha$ 
emitting galaxies also exhibit blueshifted absorption and 
Ly-$\alpha$ red wings that could arise in outflows (Frye \etal\ 2002;
Dawson \etal\ 2002; Sawicki \etal\ 2008).  
AGN also drive high-velocity outflows (e.g. Morganti \etal\ 2005;
Nesvadba \etal\ 2006; Trump \etal\ 2006); these might affect the
host galaxy ISM.  The increased frequency
of AGN and QSOs at high redshift could have a significant
effect on subsequent galaxy evolution, although the efficiency of
coupling an AGN outflow to the galactic ISM is not well understood.

In summary, at low redshift
we know that galaxies with high SFR surface density have outflows,
and at least some post-starbursts also do.  At high redshift,
galaxies selected by methods that find high SFRs show outflows,
but following the destiny of these progenitors to the present 
day is uncertain.

In this work we investigate \mgii\ absorption in star-forming
galaxies with spectra from the DEEP2 redshift survey (Davis \etal\ 2003),
which reach \mgii\ at $z\sim 1.4$.
\mgii\ absorption probes cool, 
atomic gas (neutral gas and \hii\ with metals
in low ionization states), as does Na I; \mgii\ has
an ionization potential of 15.0 eV while Na I is 5.1 eV, so \mgii\
is less easily photodissociated and
suffers smaller ionization corrections.  The \mgii\ line has 
not been used to probe galactic-scale or star-formation driven
outflows prior to 
Tremonti \etal\ (2007), mostly due to its location in the near-UV.
\mgii\ outflows are
seen in a small fraction of Seyfert nuclei (Crenshaw \etal\ 1999) and
in low-ionization broad absorption line QSOs 
(LoBALs, which are $\sim 1$\% of QSOs; Trump \etal\ 2006).
\mgii\ has 
been used in many studies of QSO absorption systems at 
intermediate redshift; these show that strong \mgii\ systems 
(EW $>0.3$ \AA) occur within 
a few tens of kpc of luminous galaxies (e.g. Lanzetta \& Bowen 1990; 
Bergeron \& Boisse 1991; Steidel, Dickinson \& Persson 1994).
Possible sources of these \mgii\ systems include extended
rotating gas, tidal debris, associated clustered absorption,
and outflows (e.g. 
Charlton \& Churchill 1998; Steidel \etal\ 2002;
Bouch{\'e} \etal\ 2006; Kaprczak \etal\ 2007; Tinker \& Chen 2008;
Wild \etal\ 2008).

Here, we use co-added spectra of 1406 galaxies
at $z =1.28$ to 1.5 from the DEEP2 survey
to detect low-ionization outflowing gas in absorption
lines of \mgii\ 2796, 2803 \AA\ and \mgi\ 2852 \AA.  
We find absorption strengths of this low-ionization gas
comparable to strong \mgii\ QSO absorption systems.
The characteristic outflow velocities are of order 300--500 \kms.
We use $K$-band and Spitzer/MIPS 24 \um\ data for part
of the sample to calibrate stellar mass and star formation
rate estimates, and explore the dependence of outflows
on mass and SFR.

We assume a LCDM cosmology with $H_0=70$, $\Omega_m=0.3$, and
$\Omega_\Lambda=0.7$.  Magnitudes in this paper are in the
AB system unless otherwise noted, and wavelengths at $>2000$ \AA\ are 
given in air, although
the reduction was done on vacuum wavelengths.

We describe the sample of galaxies and spectroscopy
in Section \ref{sec-deepspec}.  Section \ref{sec-allcoadd}
presents the composite spectrum and introduces the evidence for 
outflows in Mg II absorption, while Section \ref{sec-outflowintrinsic} 
analyzes and quantifies the outflow component of Mg II.  In Section
\ref{sec-sampleprops} we compute restframe properties 
of the host galaxies and in section \ref{sec-propdepend}
we measure the dependence of outflows on the host galaxy properties.
Section \ref{sec-discussion} analyzes the observations, including the
physical properties of the outflow and the implications for wind models,
and \ref{sec-conclusions} summarizes the conclusions.
The key results of the paper are found in Sections 
\ref{sec-allcoadd}, \ref{sec-propdepend} and \ref{sec-discussion}.

\section{Sample and Data Extraction}
\label{sec-deepspec}

\subsection{Spectroscopic data and sample selection}

The sample of galaxies we study is drawn from the DEEP2 galaxy
redshift survey (Deep Extragalactic Evolutionary Probe 2; Davis
\etal\ 2003).  DEEP2 is selected from $BRI$ imaging with the 
CFHT CFH12K camera (Coil \etal 2004b), and has obtained 
redshifts for 
approximately 32,000 galaxies, using the DEIMOS spectrograph 
at the Keck II telescope.  The DEEP2 sample is distributed
across four fields and is selected to have $18.5 < R_{AB} < 24.1$.
In three of the four fields, a color selection in $B-R,R-I$
is applied to
exclude galaxies at $z<0.7$ (Coil \etal\ 2004a), which does
not affect the high-redshift subsample we use in this paper.
The fourth field, the Extended Groth Strip, allows testing 
the color selection and confirms that the excluded galaxies
are at $z<0.7$.

DEEP2 DEIMOS spectra are taken with a 1200 lines/mm grating 
and 1.0\arcsec\ slit, yielding resolution $R \sim 5000$ and
a scale of 0.33 \AA/pixel.  A typical DEEP2 spectrum
has wavelength coverage $\sim 6600-9100$ \AA; the exact limits
vary depending on the object's location on the slit mask.
Galaxy spectra are reduced by an automated pipeline,
extracted to 1-d in a window that is typically 1.5 times the
spatial FWHM of the object along the slit, and processed
by an automated template fitter to determine a set of candidate
redshifts.  The program fits a linear combination of an artificial
emission-line template containing \oii, \hbeta, \oiii, and \halpha,
and an early type galaxy template derived
from the SDSS Luminous Red Galaxy sample (Eisenstein \etal\ 2001),
shifting the templates relative to the spectrum and computing
$\chi^2$.  It finds minima in $\chi^2$ and nominates these as
candidate redshifts.
The program also attempts to fit stellar and QSO templates.

Each spectrum is then examined by a member of the 
DEEP2 team to select the correct redshift or reject 
the redshift as unconfirmed.  To be accepted as
a confirmed redshift, a galaxy must show at least two spectral
features.  The DEIMOS spectra resolve the \oii\
$\lambda\lambda$ 3726, 3729 \AA\
doublet and a detection of both lines counts as two features.
For galaxies at $0.8<z<1.5$, \oii\ emission is the primary redshift 
indicator in DEEP2.  Beyond $z\sim 1.5$, \oii\ is increasingly 
difficult to detect in the OH airglow line forest, so the DEEP2
wavelength range was chosen to only extend as far as \oii\ at
$z=1.5$.


To study the properties of the Mg II near-UV absorption doublet at
$\lambda\lambda$ 2795.53, 2802.71 \AA, we selected DEEP2 spectra 
of confirmed galaxies that have wavelength coverage reaching
to 2788.7 \AA\ in the restframe
(--730, --1500 km/sec relative to Mg II 2795.53, 2802.71 \AA), 
or bluer.  The limited
wavelength coverage of DEEP2 spectra and the requirement of \oii\
to measure an accurate redshift mean that 
this sample is limited to galaxies at the highest redshifts in
DEEP2, $z \sim 1.4$, where \oii\ is detected and \mgii\ is in
the DEEP2 spectrum.  At $z=1.4$, Mg II
is observed at $\sim 6720$ \AA\ and \oii\ is at $\sim 8950$ \AA.
Bluer UV absorption lines 
such as Fe II or C IV cannot be measured with these data.
A small number of objects with $z>1.5$ 
also meet the wavelength coverage criterion, but objects with 
$z>1.5$ in DEEP2 are generally QSOs identified by UV emission,
so we excluded all objects with $z>1.5$.  

From a catalog with 32,308 confirmed galaxy and QSO redshifts,
these criteria select 1409 unique objects at $z<1.5$.
Three are AGN/QSOs with strong broad Mg II emission, identified in the
DEEP2 catalog as CLASS=AGN, and we exclude them.  The 1406 galaxies
are hereafter called the Mg II sample.
31 of the galaxies were observed twice, and we used only the first
observation.  All 1406 galaxies have broadband $BRI$ magnitudes
from the CFHT CFH12K observations.
The median redshift of the 1406-galaxy sample is 1.379
and the 5\%-95\% range is 1.314 -- 1.451.  
The DEEP2 survey is selected at $R$-band, so at $z\sim 1.4$ it
is effectively selected at restframe 2800 \AA.  Therefore
the sample strongly favors star-forming galaxies in the 
``blue cloud'' of the bimodal galaxy color-magnitude diagram 
(Weiner \etal\ 2005; Willmer \etal\ 2006).  

From comparing the measurements of 1453 galaxies in the
entire DEEP2 sample that were observed twice, the errors on a 
single redshift measurement are $\sigma(z) = 1.3 \times 10^{-4}$
and $5.4 \times 10^{-4}$ 
at 68\% and 95\% probability respectively.  The errors on relative 
velocity are $\sigma(v) = c\sigma(z)/(1+z)$, yielding 16 and 68 \kms\
at 68\% and 95\%, at $z=1.4$.  
In this paper, we are concerned
with the velocity of \mgii\ relative to a systemic velocity
measured at \oii.  Since these are at the extreme blue and red ends 
of the DEEP2 spectrum, the relative velocity requires a good global
wavelength calibration.  We tested this empirically by comparing
the relative velocities of \oii\ 3727 and \oiii\ 5007 for galaxies
with both lines, at $z\sim 0.8$ where they are at blue and red ends
of the DEEP2 spectrum.  In 3772 galaxies, the median 
offset is $0.9~\kms$ and the 68\% and 95\% ranges are $\pm 36$ and 
$\pm 169$ \kms\ respectively.  Outliers are caused in part by errors in
the emission line fitting due to sky residuals and weak-emission
objects, so the actual wavelength scale is better.  These $1\sigma$
redshift and wavelength accuracies are considerably smaller
than the few hundred \kms\ outflow offsets we show to be typical
in the Mg II sample.

\begin{figure}[t]
\begin{center}
\includegraphics[width=3.5truein]{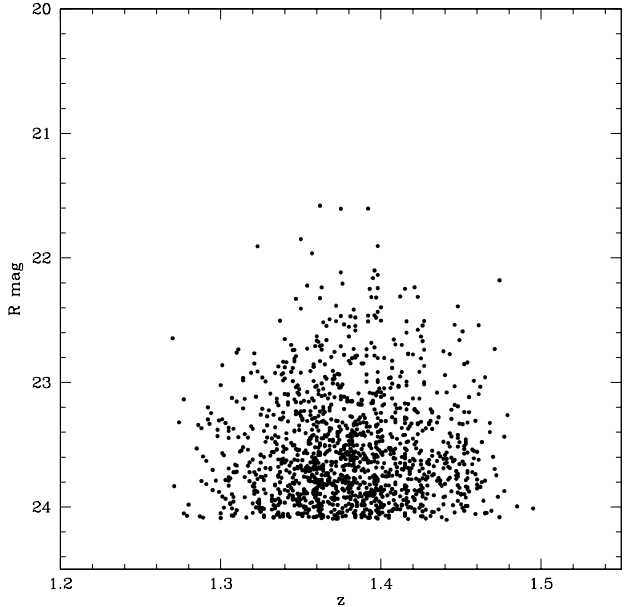}
\caption{Redshift and $R$ magnitude of the 1406 galaxies in
the \mgii\ sample.  The redshift limits are set by the wavelength
coverage of DEEP2 and $R=24.1$ is the DEEP2 magnitude
selection limit.
}
\label{fig-zdistrib}
\end{center}
\end{figure}

\subsection{Co-addition of DEEP2 spectra}
\label{sec-coaddmethod}

The median signal to noise of DEEP2 spectra in the Mg II sample
near 2820 \AA\ restframe is 0.55 per pixel, or 1.1 per FWHM.  
The typical exposure time of 1 hour
is designed for acceptable detection of nebular emission lines,
but does not yield S/N/pixel high enough to study absorption lines
in detail in most of the objects.  We construct a high S/N composite
by co-adding spectra in their rest frame.

We co-add the spectra using an IDL routine
written by members of the DEEP2 team.  From an input catalog
of DEEP2 galaxies, the routine collects their 1-D extracted spectra,
corrects each for telluric absorption, shifts the wavelength
scale to the rest frame based on the DEEP2 catalog redshift, 
which is derived from \oii\ in these objects, and coadds the 
spectra, weighting at each pixel by the inverse variance.  

Weighting solely by the inverse variance would mean that bright
objects contribute super-linearly to the coadd, since they have
higher flux and lower variance.  For the present sample, for
each spectrum we renormalized the inverse variance array
so that it has median of 1, prior to the
telluric correction.  The effect is to retain the dependence
of the inverse variance on wavelength in each individual
spectrum, e.g.\ due to night sky lines, but to take out the
object-to-object difference, to avoid weighting
bright galaxies more due to the higher S/N of their spectra.
The coadded spectrum is 
light-weighted, not equal-number-weighted, across the galaxy sample.
We did not flux-calibrate the spectra prior to coaddition; since
the galaxies occupy a narrow redshift range, the instrumental
response does not vary greatly in the pixels contributing to
a given rest wavelength.
The pixel spacing in the coadded spectrum is 13.8 \kms\ at 2800 \AA.
The S/N/pixel in the coadded spectrum of 1406 galaxies is 21,
and in the subsamples of Section \ref{sec-propdepend} the S/N/pixel
is 8 to 14.

\section{Mg II absorption in DEEP2 galaxies}
\label{sec-allcoadd}

\subsection{The coadded spectrum at $z=1.4$}

\begin{figure}[t]
\begin{center}
\includegraphics[width=2.7truein,angle=-90]{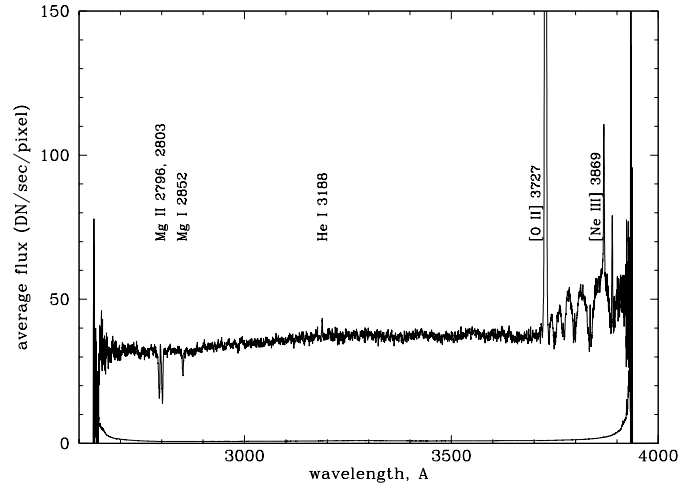}
\caption{The coadded spectrum of all 1406 galaxies in the 
\mgii\ sample, in the rest frame.  The upper line is the 
spectrum, smoothed with a 5 pixel boxcar, and the lower smooth
line is the error spectrum of the smoothed coadd.
}
\label{fig-fullspec}
\end{center}
\end{figure}

Figure \ref{fig-fullspec} shows the full coadded spectrum
of the 1406-galaxy Mg II sample.  The spectrum is dominated
by strong \oii\ 3727 emission.  There are prominent Balmer
absorption lines indicating a young stellar 
population with A stars, \neiii\ 3869 emission, and some narrow
Balmer emission.  The UV spectrum blueward of \oii\ is
relatively featureless except for weak He {\sc I} 3188 \AA\
emission, and strong absorption from \mgi\ at 2852 \AA\
and the \mgii\ doublet at 2796, 2803 \AA.

\begin{figure}[t]
\begin{center}
\includegraphics[width=3.5truein]{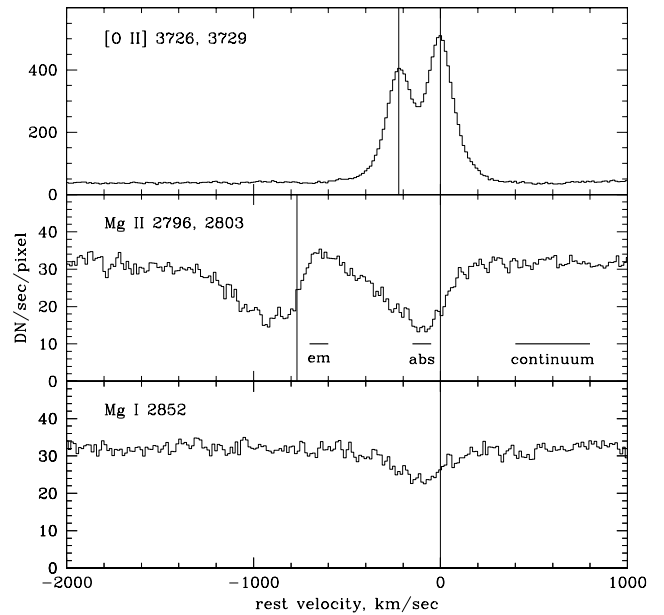}
\caption{The \oii\ 3726.0, 3728.8, \mgii\ 2795.5, 
2802.7, and \mgi\ 2852.1 \AA\ lines in the coadded spectrum 
of 1406 galaxies, relative to zero velocity as 
defined by the redshift derived from \oii.  The \oii\ doublet 
lines are at their nominal systemic velocities, but 
the \mgi\ and both \mgii\ lines show
blueshifted and asymmetric absorption profiles.
The horizontal bars in the middle panel 
show the extent of the windows used in Section \ref{sec-indivfluxes}
to measure absorption and excess emission in individual spectra.
}
\label{fig-stackspec}
\end{center}
\end{figure}

In Figure \ref{fig-stackspec} we extract the regions of the
co-added spectrum around the lines \oii\ 3726.0, 3728.8 \AA, 
\mgii\ 2795.5, 2802.7 \AA, and \mgi\ 2852.1 \AA, and plot 
them on restframe velocity
scales derived from the DEEP2 catalog redshift.  At this high
redshift, the DEEP2 redshift is based solely on fitting \oii.
The \oii\ doublet is thus necessarily fixed to zero velocity,
here referenced to the redder line at 3728.82 \AA.
The immediately visible result is that the \mgii\ and \mgi\ 
absorption lines are asymmetric and blueshifted by a few
hundred \kms\ with respect to the systemic velocity defined
by \oii.

The lines of the \oii\ doublet are symmetric, with low-intensity
wings more extended than Gaussian.  The wings could be due to wings
in individual galaxies or simply the fact that the summation
of Gaussians of different dispersions is not itself a Gaussian.
In local superwind galaxies, faint optical emission is 
seen from high-velocity wings of a few hundred \kms,
from shocked gas swept up by the wind
(e.g. Bland \& Tully 1988; Heckman \etal\ 1990).
The \neiii\ line, which arises in AGN and in very blue starforming 
galaxies, is much weaker than \oii, and has a small blueshift
of 40 \kms.

In local luminous infrared and starburst galaxies, the 
centroid of optical emission can be slightly blueshifted with
respect to the systemic velocity, presumably due to emission 
from outflowing gas.  For IR-luminous
galaxies, Mirabel \& Sanders (1988), found the optical 
catalog redshift was blueshifted by 90 \kms\ on
average relative to the centroid of 21 cm \hi\ emission.
In local edge-on starbursts, Lehnert \& Heckman (1996)
found the nuclear emission lines blueshifted by a median of
25 \kms\ relative to the overall rotation curve.
However, in our spectra, since the galaxies are comparable in size to
the slit width, \oii\ emission is integrated over most of the 
galaxy.
In our coadded spectrum, the velocity of \oii\ and the high order 
Balmer lines at 3798, 3835, and 3889 \AA\ agree well.  However,
centroiding the broad Balmer stellar absorption lines is difficult due to
narrow Balmer emission.  We coadded spectra of all DEEP2 blue galaxies
at $1.1<z<1.5$ to reach the Ca II 3933.7 stellar absorption line,
and found a negligible offset of $<5~\kms$ between Ca II and 
\oii.\footnote{Outflows are rarely seen in Ca II H and K lines in DEEP2, 
due to stronger stellar absorption, lower abundance than Mg, 
greater depletion onto dust, and a large ionization correction 
for Ca II.  Similar factors make Ca II rare in QSO absorption line
systems (Bowen \etal\ 1991; Wild \etal\ 2006).}
Thus the \oii-derived redshift is a reliable measure of the systemic
velocity.

The \mgii\ absorption profiles in the middle panel of
Figure \ref{fig-stackspec}
are strikingly offset from the systemic velocity.  
Both lines of the \mgii\ doublet
show an asymmetric, blueshifted absorption trough.
There is a small amount of absorption redward of systemic
to $+100$ \kms,
a deep trough reaching $\sim 60\%$ absorption at a
blueshift of --100 \kms, and a asymmetric, triangular
``sawtooth'' absorption profile
extending blueward beyond $-500$ \kms.  The \mgi\ line
shows a similar profile, although the depth of absorption is weaker.
In the coadded spectrum of the 1406-galaxy sample, the
equivalent width of \mgii\ 2796+2803 absorption is 4.6 \AA, and
of \mgi\ 2853 absorption is 0.9 \AA.

Between the two \mgii\ lines, there is a hint that the 
spectrum goes higher than continuum, at $\sim -600$ \kms,
or $\sim +170$ \kms\ relative to the Mg II 2796 line.
We study this signature in more detail in Section 
\ref{sec-excessem}, where it is shown to be strong in 
a small fraction of the sample and due to
Mg II emission, most probably from narrow-line AGN.

The velocity offset of \mgii\ is far greater than the 
uncertainties in the DEEP2 wavelength scale, as discussed in 
Section \ref{sec-deepspec}.  Additionally, the 
velocity spread of the absorption is greater than the
velocity width of the galaxy emission lines, and 
the asymmetric absorption profile could not arise from 
calibration errors.  Both the blueshift and the 
asymmetric profile are real properties of the sum of
UV emission over the 1406 galaxies.

\subsection{The cause of blueshifted absorption}

The strong blueshifted absorption shown in Figure \ref{fig-stackspec}
is the chief observational result of this paper.  It indicates
the presence of a significant amount of gas in front of and
flowing away from the continuum source at several hundred \kms\ in 
both \mgii\ and \mgi, which are low-ionization lines that arise 
in cool gas.  Here we discuss the physical interpretation of 
this outflow.

The DEEP2 $z\sim 1.4$ galaxies are small in apparent size; in
the subset with HST imaging, 95\% have $r_{eff}<0.95\arcsec$.
They are observed with a 1\arcsec\ slit and the seeing
smooths over the galaxy light distribution, so that the 
continuum in the spectra is the integrated light of the galaxy.
Because the absorption reaches $>50\%$ depth, the
absorbing material must be spread over the galaxy, rather than,
for example, being confined to a nucleus that does not dominate the
integrated light.

\subsubsection{Circumstellar versus galactic-scale absorption}

First we ask whether the absorption is galactic-scale or
circumstellar.
Hot stars often show ionized outflows in high-ionization lines
such as C IV, but do not show outflows in low-ionization lines.
Stellar \mgii\ absorption is weak in luminous, hot O and B
stars (Lamers \etal\ 1973; Lamers \& Snijders 1975; Kondo \etal\ 1975).
Strong and sometimes blueshifted absorption is found in A type
supergiants (Praderie \etal\ 1980; Verdugo \etal\ 1999).
The very brightest and rarest
late B, A, and F stars can have asymmetric or blueshifted 
absorption of order -100 km/s, but fainter stars do not (Snow \etal\ 1994).
Outflows in Mg II are sufficiently infrequent that
they should not dominate the integrated light of a star-forming
stellar population, and the known stellar outflows are lower velocity
than observed in the stacked galaxy spectrum.

The firmest evidence that the outflows seen in DEEP2 are
galactic-scale rather than circumstellar comes from the 
similarity of the \mgii\ and \mgi\ outflow profiles.
\mgi\ 2852 is absent in hot stars,
present in F stars, but is photospheric and
does not show signatures of mass loss (Snow \etal\ 1994).  
Circumstellar outflows cannot contribute to the outflow signature in \mgi.

Photospheric, non-outflowing \mgii\ absorption is weak in very
blue stars and stronger in cooler stars (e.g. Rodriguez-Merino
\etal\ 2005).  Some of the zero or low-velocity
\mgii\ absorption could be stellar, but this cannot account for
the high-velocity blueshifted absorption.

\subsubsection{Clustering of intervening absorbers}

Intervening \mgii\ absorption line systems seen against QSOs
are known to be associated with normal, relatively luminous 
galaxies, out to impact parameter $\sim 50$ kpc, and to be
clustered similarly to luminous galaxies (e.g. 
Steidel \& Sargent 1992; Steidel \etal\ 1994; 
Bouch{'e} \etal 2006; Wild \etal\ 2008).
``Strong'' \mgii\ absorbers (EW$ >0.3$\AA)
have average EW of order $W_{2796} \sim 1 \AA$
and abundance $dN/dz \sim 1$ (Nestor \etal\ 2005).
One could imagine that absorbers that are not directly
associated with the source galaxy, but clustered around it,
would produce a blueshifted absorption signature, tailing
off to more negative velocities.  In this picture the 
absorption velocity would not represent an outflow but 
simply the location of the clustered absorbers in the
Hubble flow.

However, the absorption strength that could plausibly be
produced by nearby clustered absorbers is much lower than we observe.
There are a variety of ways of estimating this, which rely
on the frequency of intervening absorbers $dN/dz$, their
similar clustering to luminous galaxies, and/or that they are 
associated with luminous galaxies.  

For example, consider
absorbers clustered around galaxies with $\xi(r) = (r/r_0)^{-\gamma}$,
$r_0 = 4h^{-1}$ Mpc comoving, $\gamma=1.8$, similar to galaxies
and to absorbers clustered around QSOs
(Steidel \etal\ 1994; Coil \etal\ 2004a; Bouch{\'e} \etal\ 2006;
Wild \etal\ 2008).
At z=1.4, $r_0$ corresponds to $\Delta z = 0.0029$, or 870 \kms\ in our
spectrum.  Thus absorbers within $r_0$ could contribute to the
observed absorption.  From integrating $\xi(r)$, the overdensity within
$r_0$ is $\delta=10.5$.  The expected strength of the absorption
is roughly $W_{clust} = \delta\ \Delta z \ dN/dz <W_{absorber}>$.
From the survey of Nestor \etal\ (2005), the frequency of 
absorbers with $W_{rest,2796}>0.3$ \AA at $z=1.4$ is $dN/dz=1.22$,
and integrating over the equivalent width distribution 
yields the mean $<W_{absorber}>=1.07$ \AA.  The predicted absorption EW from
clustering is then $W_{clust} = 0.04$ \AA, a factor of $\sim 40$
fewer than we observe (Table \ref{table-absprops}).
Essentially, the overdensity within $r_0$ is not enough to
overcome the short pathlength within $r_0$, since $dN/dz \sim 1$.

Another way of looking at this problem is that strong 
intervening \mgii\ absorbers
are associated with luminous galaxies or QSO hosts with 
impact parameters $\sim 50$ kpc 
(e.g. Steidel \etal\ 1994; Hennawi \etal\ 2006; Chen \& Tinker 2008) 
with a high, but not 100\%, covering factor.  The absorption EW
we observe is $\sim 1.5\times$ the mean of strong \mgii\ absorbers, 
$<W_{absorber}>$.  To produce the absorption we observe by intervening galaxies,
every source galaxy would have to have $\sim 1.5$ interveners,
with 100\% covering factors, projected at
$<50$ kpc impact parameter and in front by $\lesssim 800$ \kms.  This 
radius and velocity separation is fairly close to the definition of 
a close pair, but the fraction of galaxies in close pairs in the DEEP2 
sample is only 10\% (Lin \etal\ 2004).

\subsubsection{Absorption depth and covering factor}

Because the Mg II absorption in Figure \ref{fig-stackspec}
is deep, 55\% absorption at the bottom of the trough, 
we know that it covers a large fraction of the UV light over the whole
sample of galaxies.  It may cover $\sim 55\%$ of the stars in all
the galaxies, or all the stars in $\sim 55\%$ of the galaxies, or some
compromise.  This also helps to confirm that it is from a galactic-scale
outflow, rather than from circumstellar outflows 
associated with a particular type of star, or outflows that are
local to an active nucleus.

Winds in local starburst disks are more
commonly detected in face-on galaxies (Heckman \etal\ 2000).
The frequency of wind detection in Na I absorption
suggests that the angular covering fraction $C_\Omega$ of their winds,
due to the wind opening angle,
is about 0.4-0.5 in starbursts and LIRGs, and higher in 
ULIRGs (Rupke \etal 2005b).  For LIRGs and
ULIRGs that do have detected winds, Rupke \etal\ found a 
clumpiness covering fraction 
within the wind of $C_f \sim 0.4$.  For a coadded sample of galaxies
that all have optically thick winds, the absorption depth is $C_\Omega C_f$,
which would be $\sim 0.2$ in the Rupke \etal\ sample.
An absorption trough depth of 55\% is thus
quite large compared to the local values, suggesting
that winds are actually happening in all of the DEEP2 
sample galaxies and that the covering fractions are higher than
observed locally in Na I.  

Additionally, the \mgi\ outflow
line is not as deep as the \mgii\ outflow, even though \mgi\
has a higher oscillator strength and \mgii\ is saturated.
The most likely explanation is that \mgi\  has a lower
covering fraction, and presumably traces  denser or cooler 
parts of the outflow.

Although the blueshifted absorption is from interstellar gas,
it is possible that some of the zero-velocity absorption is
due to stellar photospheric lines.  Separating stellar
and ISM contributions typically requires modeling the spectra 
of star-forming populations with a code such as
Starburst99 (Leitherer \etal\ 1999), but we cannot use it to predict the
\mgii\ line profiles because Starburst99 does not produce high resolution
spectra at 2800 \AA.  Currently, Starburst99, PEGASE 
(Fioc \& Rocca-Volmerange 1997), and the models of Bruzual \& Charlot (2003)
are all limited to resolutions $\sim 10$ \AA\ at 2800 \AA, which
limits our ability to model the \mgii\ line profiles or EW expected
from given stellar populations.  However, we can decompose the
absorption into velocity components to separate the outflow
and zero-velocity, non-outflowing components 
(Section \ref{sec-outflowintrinsic}).

\subsubsection{The possibility of AGN-driven outflows}

Active galactic nuclei are known to drive winds and 
to sometimes show blueshifted absorption outflow signatures
(e.g. Crenshaw \etal\ 1999; Veilleux \etal\ 2005; 
Trump \etal\ 2006; Wild \etal\ 2008).  AGN outflows in \mgii\ are seen
in low-ionization BAL QSOs, which are a few percent of all QSOs 
(Trump \etal\ 2006), and in 1 of 17 Seyfert nuclei studied by 
Crenshaw \etal\ (1999).
AGN/starburst composite IR-luminous galaxies often show low-ionization
outflows traced by Na I (Rupke \etal\ 2005c).  There are signs 
that the Na I outflows in AGN/starburst composites are different 
on average from those in starburst IR-luminous galaxies,
with lower opening angle and higher velocity.  However,
the AGN does not dominate the large scale outflow in these
galaxies (Rupke \etal\ 2005c).

The high frequency and covering fraction of absorption in the DEEP2
galaxy sample suggests 
that the outflows are driven by star formation rather than AGN.  
Nearly all blue cloud galaxies are star-forming, while luminous
AGN are much rarer than 50\% and are not common in blue
star-forming galaxies at $z\sim 0.8-1$, as measured by
X-ray or emission line signatures; AGN signatures are more
often seen in red or intermediate-color galaxies (Nandra \etal\ 2007;
Weiner \etal\ 2007).  Further, \mgii\ absorption occurs in just a few
percent of AGN, as described above.  Some of the galaxies in
our sample may have AGN producing \mgii\ emission, discussed
in Section \ref{sec-excessem}, but this does not appear to 
be correlated with the galactic-scale \mgii\ absorption.

\subsection{Absorption detected in individual spectra}
\label{sec-indivplots}

\begin{figure*}[t]
\begin{center}
\includegraphics[width=5.9truein]{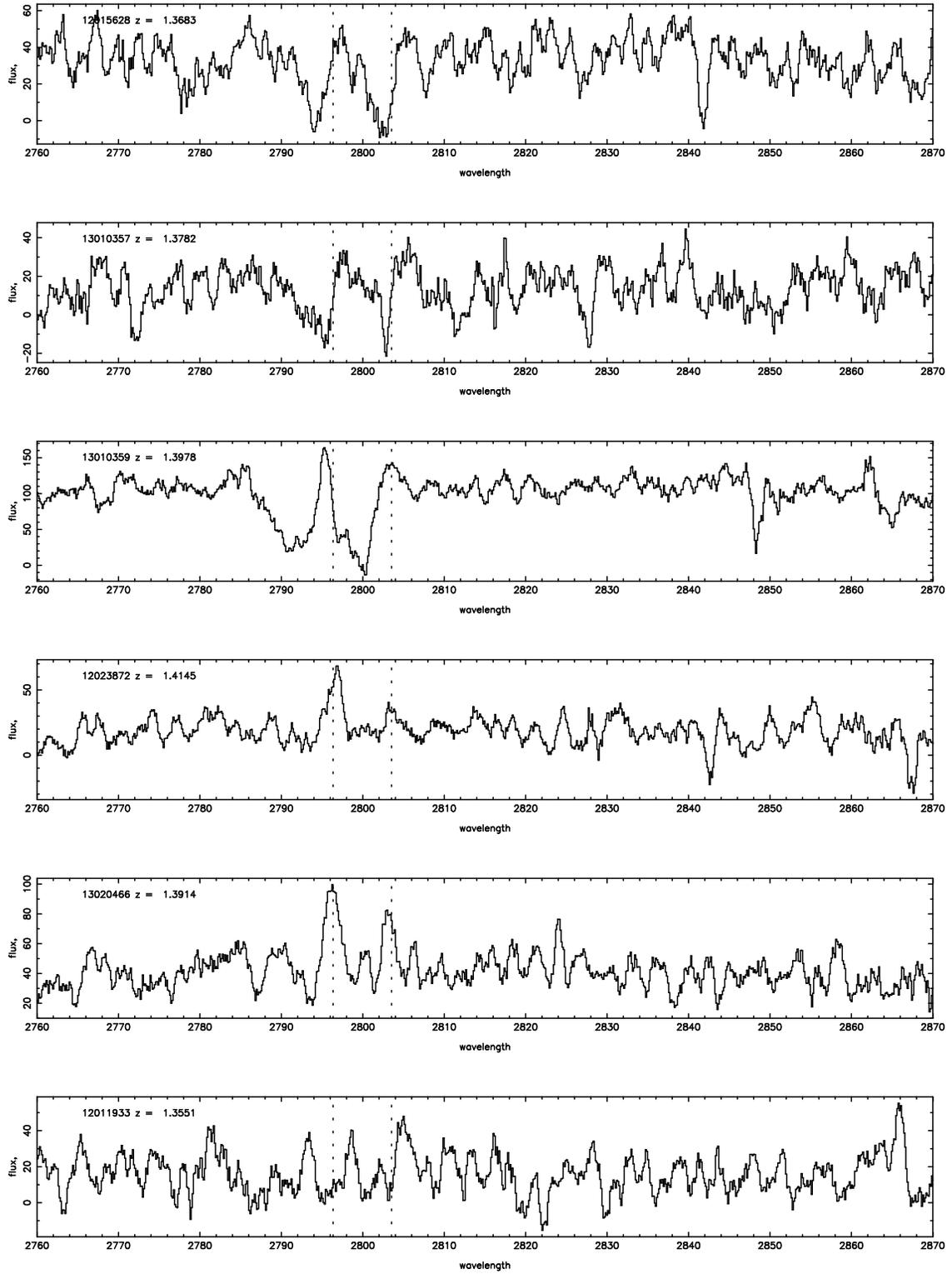}
\caption{DEIMOS spectra of six galaxies in the \mgii\ sample
in the rest frame.  The spectra have been smoothed by a 9 pixel
boxcar (1.25 \AA\ in the rest frame).  Vertical dotted lines
indicate the systemic velocity of \mgii\ $\lambda\lambda$ 2796, 2803.
The top three illustrate the $\sim 5-10\%$ of the sample
that clearly show blueshifted absorption in individual spectra,
with a broad asymmetric profile. 
Spectra 4 and 5 show narrow \mgii\ emission at the systemic
velocity, visible in $\sim 5\%$ of the spectra, presumably due to 
narrow-line AGN.  The spectrum in the bottom panel is more typical of
the majority of the sample, where \mgii\ is neither individually
detected nor ruled out.  The units of the spectra are e-/hour and the
typical noise is 40-50 e-/hour/pixel before smoothing.}
\label{fig-samplespec}
\end{center}
\end{figure*}

We examined the individual spectra in the 1406-galaxy sample
at the location of \mgii.  Figure \ref{fig-samplespec}
shows six illustrative spectra.  The top five panels have 
relatively high S/N; the top three panels show
blueshifted \mgii\ absorption, and panels 4 and 5 show narrow
\mgii\ emission.  Many
of the spectra do not have a high S/N in the continuum
and we neither detect nor rule out \mgii\ absorption.  The
lowest panel of Figure \ref{fig-samplespec} shows one 
such typical spectrum.  As discussed in Section \ref{sec-indivfluxes},
in the aggregate, these spectra do show \mgii\ absorption.

About 5-10\% of the individual spectra are high enough S/N to show
convincing \mgii\ absorption.  When it is seen, it is 
almost always blueshifted.  The top three panels of 
Figure \ref{fig-samplespec} show examples of blueshifted
absorption; the third object has especially strong absorption.
Note that the individual galaxies have asymmetric sawtooth
absorption profiles just as the coadded spectrum in 
Figure \ref{fig-stackspec} does - there is a long tail to 
high velocities.  The sawtooth profile is a
property of individual winds, not just caused by the coadding
process.

About 4\% of the individual spectra show fairly narrow \mgii\ 
emission lines, as seen in panels 4 and 5 of Figure \ref{fig-samplespec}.
This is not characteristic of star formation and probably 
comes from Seyfert nuclei or other narrow-line AGN.  
The \mgii\ emission is generally
at the systemic velocity.  The emission is sometimes accompanied
by detectable blueshifted absorption, but less than 50\% of
emission objects have individually detectable absorption.
In Section \ref{sec-excessem} we discuss the properties
of these excess-emission objects.

\subsection{Flux decrements in individual galaxies}
\label{sec-indivfluxes}

In any coadded spectrum or stacked image detection, one has
to ask if a feature is a general property of the sample or
is caused by a small fraction of outlier objects,
and if the feature is weighted more or less to brighter
objects.  This is a less severe problem for absorption
than for emission lines, but is still of interest.

Each individual DEEP2 galaxy spectrum has low S/N/pixel
at 2800 \AA.  To address the effect of stacking, for each spectrum
we compute the average pixel value and its error estimate
in three windows
relative to the Mg II 2802.7 line: (1) +400 to +800 \kms, (2)
--150 to --50 \kms, and (3) --700 to --600 \kms.  Window 1 
measures the continuum level, window 2 is located at the 
deepest part of the blueshifted Mg II 2802.7 absorption,
and window 3 is at the high point between the two Mg II 
lines, where there may be excess emission above continuum.
The locations of these windows are shown in the middle panel
of Figure \ref{fig-stackspec}.

\begin{figure}[t]
\begin{center}
\includegraphics[width=3.5truein]{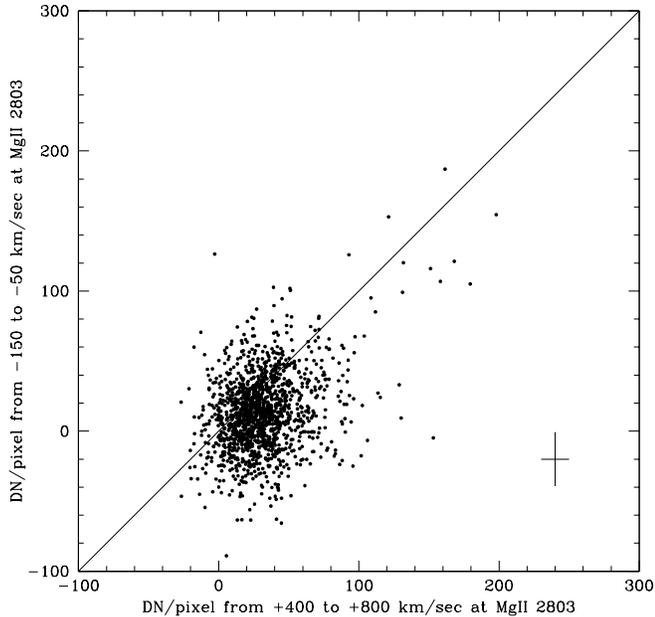}
\caption{Flux decrement blueward of \mgii\ 2803 in individual galaxies.
The error cross shows the median error bar.  The whole distribution
is shifted, with the flux
density at -150 to -50 \kms\ systematically below the continuum.
}
\label{fig-indiv-windabs}
\end{center}
\end{figure}

\begin{figure}[t]
\begin{center}
\includegraphics[width=3.5truein]{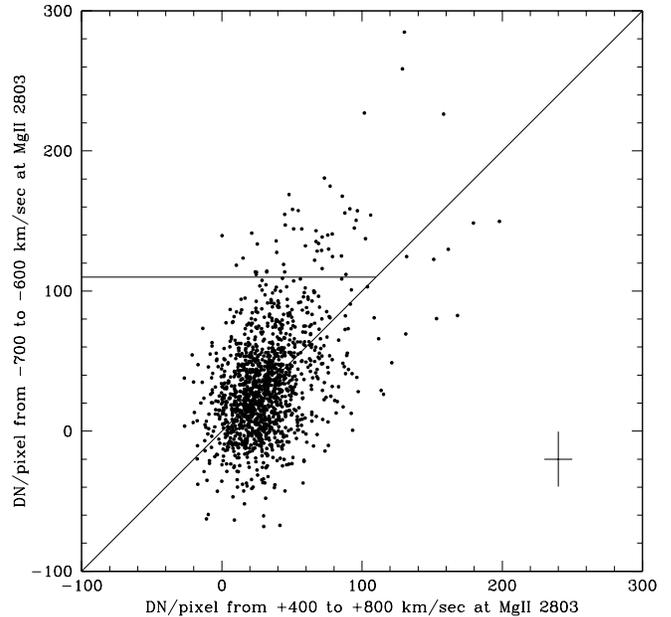}
\caption{Flux increment between the two \mgii\ absorption lines
in individual galaxies. A tail of objects have enhanced emission
above continuum in the window at --700 to --600 \kms\ relative to 
\mgii\ 2802.7, or +69 to +169 relative to \mgii\ 2795.5 \AA.
The error cross shows the median error bar.  The horizontal
line at 110 DN/pixel defines the ``excess emission'' galaxies.
}
\label{fig-indiv-pcyg}
\end{center}
\end{figure}

Figure \ref{fig-indiv-windabs} compares the average flux
in the continuum window 1 and the absorption window 2.
There is considerable scatter, as expected given the median
errorbar, but the entire distribution of points is 
shifted below the 1:1 line.  The blueshifted absorption
is a phenomenon that occurs throughout the $z=1.4$ galaxy sample,
not in just a small subset.

Figure \ref{fig-indiv-pcyg} compares the average flux
in the continuum window 1 and the excess emission window 3,
between the Mg II absorption lines.
The behavior here is rather different than in 
Figure \ref{fig-indiv-windabs}.  The bulk of galaxies
scatter around the 1:1 line; they are roughly at the
continuum level in window 3.  However, a small number
of galaxies show significant excess emission in window 3.
These are relatively bright and so contribute to the light-weighted
coadded spectrum.  We discuss these unusual objects in the
next section.

\subsection{Galaxies with excess Mg II emission}
\label{sec-excessem}

\begin{figure}[t]
\begin{center}
\includegraphics[width=3.5truein]{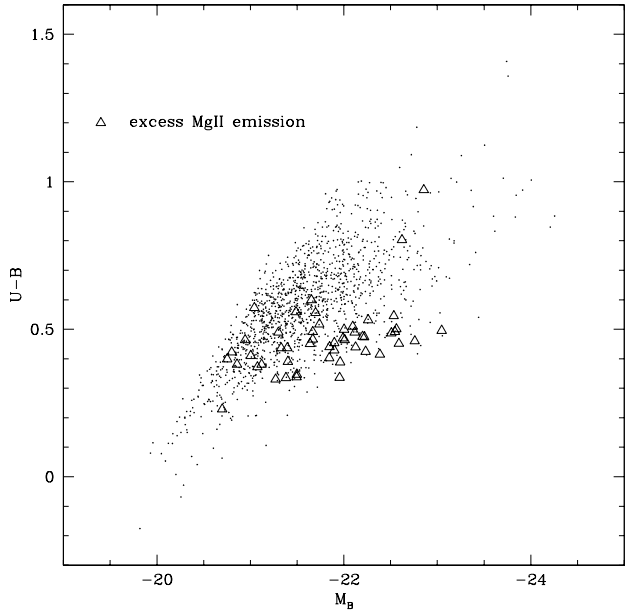}
\caption{The magnitude-color distribution of the \mgii\ 1496-galaxy
sample in restframe $M_B$ and $U-B$.  Galaxies with excess
\mgii\ emission above continuum are plotted as large open triangles.
The emission is likely from narrow-line AGN and the objects are
among the brighter of blue galaxies.
}
\label{fig-colormagexcess}
\end{center}
\end{figure}

We define the ``excess emission sample'' as the
tail of 50 objects with flux(window 3)$>110$ DN/pixel
and flux(window 3)$>$flux(window 1).  These galaxies tend
to be in the bluer half of the sample and brighter than
average, as shown in the magnitude-color diagram of
Figure \ref{fig-colormagexcess}.

\begin{figure}[t]
\begin{center}
\includegraphics[width=3.5truein]{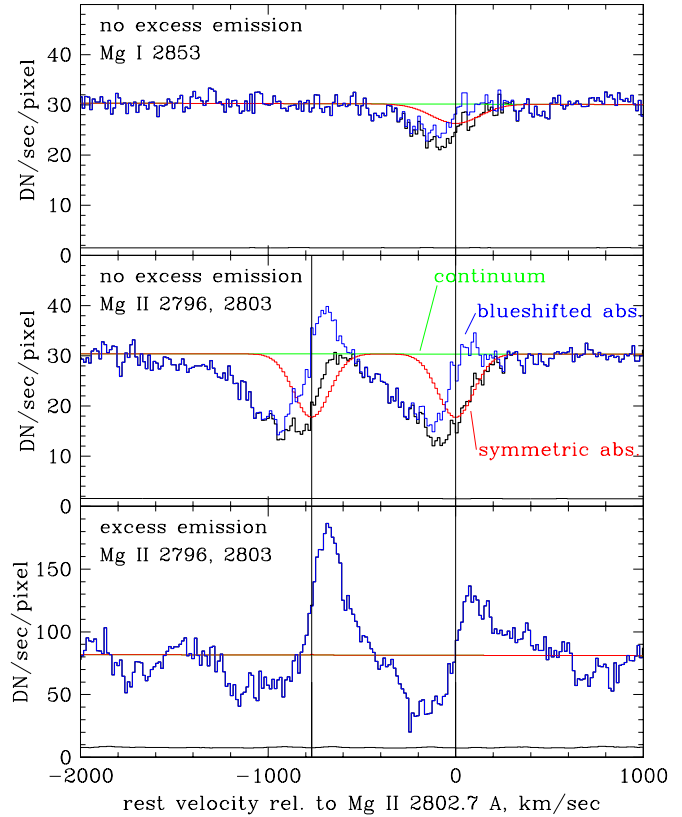}
\caption{
Absorption profiles for: 
\mgi\ 2852 \AA\ for 1356 galaxies without excess \mgii\ emission
(upper panel);
\mgii\ 2796, 2803 \AA\ for the same 1356 galaxies (middle panel); and
50 galaxies with excess \mgii\ emission between the two absorption lines 
(lower panel).  The colored lines in the middle and upper panels
show decompositions into continuum (green straight line),
symmetric absorption (red gaussians) and outflow absorption (blue),
discussed below in Section \ref{sec-outflowintrinsic}.  The
horizontal black line near zero is the error array
for the coadded spectrum.
}
\label{fig-mgemsplit}
\end{center}
\end{figure}

Figure \ref{fig-mgemsplit}
shows the coadded spectra of the 1356 galaxies without excess
emission
and the 50 galaxies with it (lower panel).
When these 50 galaxies are coadded, they show a clear
signature of emission around +100 \kms\ relative to 2795.5 \AA\
and at +100 \kms\ relative to 2802.7 \AA.
In the ``no-excess emission'' sample (middle panel of 
Figure \ref{fig-mgemsplit}), the spectrum as observed
does not have an excess above continuum, but modeling
the red wings of the two \mgii\ lines suggests a small
excess, discussed further in Section \ref{sec-outflowintrinsic}.

On examining the individual spectra in the excess-emission galaxies,
a number of them show relatively weak \mgii\ emission lines with 
FWHM $\sim 150$ to 300 \kms, in which the doublet is resolved.  
This \mgii\ emission is weaker and narrower than the broad
\mgii\ of QSOs, and is probably from Seyfert nuclei (e.g.
Wu, Boggess \& Gull 1983).  Another possibility is
redshifted emission from the back side of the wind;
this has been observed locally
in the Na I line in the galactic wind of NGC 1808 (Phillips 1993).
Chromospheric \mgii\ emission from stars
is unlikely to be the cause, especially since it is less strong
in hot stars (e.g. Snow \etal\ 1994), and the emission we
see is stronger in bluer galaxies.

Some of the excess-emission spectra also
show \mgii\ absorption, near systemic velocity or blueshifted 
by a few hundred \kms.  At first glance the composite 
spectrum in the lower panel of Figure \ref {fig-mgemsplit} resembles
the redshifted emission in P Cygni absorption/emission profiles, but
the emission in individual spectra is generally near the systemic
velocity.  Many objects in the excess-emission sample do not
have a P Cygni profile, but in the coadded spectrum, the 
overall blueshifted absorption eats away at the Seyfert emission lines.

In the lower panel of Figure \ref{fig-mgemsplit}, the emission 
peak of the \mgii\ 2795.5 line is higher, and the continuum
is not flat, showing ``shoulders'' out to $\sim 600$ \kms\ away
from the \mgii\ lines.  The bluer \mgii\ line has a higher $g$-factor
and should be stronger.  More exotically,
it is also possible that the line
contains some weak broad \mgii\ AGN emission,
which is then partially absorbed by blueshifted gas, leaving
a trace of asymmetry and the broad continuum shoulders.
We searched for Ne V AGN emission at 3425 \AA\ but did not detect any
in coadds of either the excess or no-excess spectra.

If the \mgii\ emission is from AGN, it is not clear why it
is strongest in the bluest and lower-mass subsample.  Some AGN
light might be contributing to the galaxy color and luminosity.
The emission might also be redshifted 
from the far, receding side of the wind, as in P Cygni profiles.
Blueshifted absorption and redshifted emission is seen in the
Na I D line in the wind of the star-forming galaxy NGC 1808 
(Phillips 1993), where the wind is found across the disk and
is due to star formation rather than nuclear activity.
However, it is not clear why back side emission would be
stronger in the bluer galaxies.  Bluer, lower-mass galaxies
might have less internal extinction so that it is easier to
see the back side of the wind.  But this does not appear
to explain the individual \mgii\
lines that are at the systemic velocity.
Interpretations of the emission
are hampered by the lack of moderate-resolution \mgii\ spectra
in local galaxies to use as a comparison.

The excess-emission sample appears to contain relatively faint
narrow-line AGN, some detected in individual spectra
in Mg II emission, and possibly low-level broad-line AGN, or
back-side wind emission.  However, the strength of blue-shifted
absorption is similar between the composite of this sample and of 
the remaining 1356 galaxies: the outflow component seems to be 
independent of the presence of the AGN.  

Because the excess-emission galaxies have a distinctly
different spectral shape and the underlying emission makes it
harder to measure the properties of the absorption,
we exclude them from the discussion of Section \ref{sec-propdepend}
where we measure the dependence of the absorption on galaxy 
properties.  When the remaining 1356 galaxies are analyzed, we
do see some residual emission, especially in the bluer galaxies;
see Section \ref{sec-outflowintrinsic}.

\section{Intrinsic and outflow absorption}
\label{sec-outflowintrinsic}

The spectra shown in Figure \ref{fig-stackspec}
show \mgi\ and \mgii\ absorption that is asymmetric but with
both blue and red-shifted absorption.  We propose 
a simple model to decompose the absorption into
two physically meaningful components: intrinsic
and outflow.

\subsection{Decomposition into symmetric and blueshifted absorption}

In composite spectra, we cannot hope to resolve individual
absorption systems.  We aim to differentiate intrinsic
absorption within the galaxy at the systemic velocity 
from outflow absorption at negative velocities.  The
task is complicated by the fact that \mgii\ is a doublet.
Our approach is to measure the absorption at systemic
velocities from the red side of the redder \mgii\ line
and the outflow in the blue side of the bluer line.
A main goal of this decomposition is to avoid counting
intrinsic absorption in the wind opacity; the results on
outflows are not very sensitive to the details of the
method.

Suppose that a spectrum is produced by a galaxy
with a mix of UV-emitting stars and gas with some
velocity dispersion or rotation, and surrounded by a wind
of outflowing material.  
The actual line-of-sight velocity
of the galactic ISM absorption will depend on the geometry
of the galaxy, but when a large number of spectra are
coadded, this should average out, so that 
absorption due to galactic gas will have roughly 
zero mean velocity and the average velocity spread
of the coadded galaxies.  Stellar photospheric absorption
will have a zero mean velocity and width governed by the
stellar atmospheres.
The intrinsic ISM+stars component should be symmetric about
zero velocity, and its red wing can produce the 
positive-velocity absorption.  

In principle, inflow of gas accreting onto the galaxy 
could also contribute to the red wing.  However, infalling
gas whose kinematics are governed by gravitation will be at
velocities of order the virial velocity dispersion of the halo,
and it is very difficult, even if gas is detected, to
separate inflow from any other
gas associated with the galaxy.  In contrast,
outflowing gas with kinematics set by star formation or AGN
can be distinguished.

Outflowing gas that lies between the UV source and
us will absorb at blueshifted, negative velocities.
This gas could have some velocity dispersion, but it should
be fairly small in cool gas (e.g. Rupke \etal\ 2005a).
To determine properties of the outflowing component,
we attempt to separate it from the intrinsic component.
We adopt the model:

\begin{equation}
F_{obs}(\lambda) = C F_{em} (1-A_{sym}) (1-A_{flow}),
\end{equation}

\begin{equation}
A_{sym}(\lambda) = A_1 G(v,\lambda_1,\sigma) + 
A_2 G(v,\lambda_2,\sigma)
\end{equation}

\noindent
where $F_{obs}(\lambda)$ is the observed flux density, $C(\lambda)$ is the 
underlying linear fit to the continuum of the galaxy, $F_{em}(\lambda)$ 
describes any variation in the emission beyond the 
linear continuum slope, and $A_{sym}$ and $A_{flow}$
are the depths of the intrinsic (symmetric) and outflow (blueshifted)
absorption.  We model the symmetric absorption as the product
of two Gaussians $G(v)$ centered on the wavelengths $\lambda_1,\lambda_2$
of the two \mgii\
lines, with velocity dispersion $\sigma$ and intensities $A_1,A_2$.

The result of this model is illustrated in the middle
panel of Figure \ref{fig-mgemsplit}.  The heavy black line is the
observed spectrum; the green straight line is the continuum fit,
the red double Gaussian absorption lines are
the symmetric component $A_{sym} = A_1 G(2796) + A_2 G(2803)$, and the 
blue line is the outflow component $1-A_{flow}$ that remains after the
observed spectrum is multiplied back by $1-A_{sym}$.  

The relative intensities of the absorption in the two lines
depend on the balance between optical depth and covering 
factor.\footnote{For individual
absorption systems, one must also consider the possibility
that spectrograph resolution dilutes unresolved but saturated lines.
Our measurements are of an ensemble of coadded absorbers,
and the ensemble is resolved though individual clouds are not.}
For the \mgii\ doublet, the $g$-value of the bluer line
and its oscillator strength are twice as large,
so in optically thin absorption the bluer line is twice
as strong, while in optically thick absorption the strengths
are equal.  The \mgii\ transition becomes optically 
thick at a low column density ($N_{Mg II} \gtrsim 10^{13}$
atoms cm$^{-2}$, or $N_H > 10^{18}$ for solar abundance), 
so it is generally saturated in cool gas and in stars.
In our spectra the depths of the \mgii\ lines appear equal, so we
set $A_1=A_2$, implying saturated absorption in the symmetric 
component.

Decomposing the symmetric and blueshifted components 
can be awkward due to the doublet.  We infer the properties
of the symmetric components from the red side of the red line.
By assuming symmetry, we can determine how much of the
absorption at negative velocities is actually due to the symmetric
component.  We assume the strengths of the two \mgii\ lines
are equal to infer the symmetric component of the
bluer line.

The upper two panels 
of Figure \ref{fig-mgemsplit} illustrate the decomposition applied
to the \mgii\ and \mgi\ 2852 lines.  The x-axis is 
velocity referenced to the $\lambda=2802.7$ or the $\lambda=2852.1$ line.
We first fit the linear continuum $C(v)$ by averaging the data in
two windows at $-2400<v<-1600$ \kms\ and $800<v<1600$ \kms,
and interpolating between those points.  We divide
the observed spectrum by the continuum so that the line strengths
are fitted to a spectrum normalized to 1.

We then fit a Gaussian line profile $A_2 G(v,\lambda_0=2802.71,\sigma)$ to 
the positive velocity data of $1-F_{obs}/C$, using only the pixels 
at $0<v<1600$ \kms.
The fitted parameters are a constant continuum level,
the intensity $A_2$, and the dispersion $\sigma$; the central
velocity is held fixed at 0.  The parameters are tabulated in
Table \ref{table-absprops}.  The dispersion is due to both 
velocity dispersion and the intrinsic widths of stellar \mgii\
lines, which can be large for cooler stars.  The fit is performed using
a Levenberg-Marquardt non-linear least squares minimization.
Setting $A_1=A_2$, we duplicate this Gaussian for the 
bluer \mgii\ line to make $A_{sym}(v)$.

Knowing $F_{obs}(v)$, $C(v)$, and $A_{sym}(v)$, we 
obtain $F_{em}(v) \times (1-A_{flow}(v))$.  If the galaxy
emission is featureless, then $F_{em}=1$ and we have measured
the absorption of the outflow component $A_{flow}$.  If
there is some residual \mgii\ emission,
then $F_{em}(v)$ can be $>1$ and we may underestimate the
depth of the outflow absorption.  This is a fairly small effect
once the 50 excess-emission galaxies have been removed
from the sample, but reappears in the subsample of the bluest
galaxies (Section \ref{sec-propdepend}).

Ultimately, the details of decomposing the symmetric absorption
do not have a strong effect on the interpretation of the
wind component.  We attempt this decomposition to keep from 
erroneously counting too much symmetric absorption in the wind; 
this is important since some of the symmetric absorption in
both \mgii\ and \mgi\ could be
due to photospheric or zero-velocity ISM rather than outflow absorption, 
especially in galaxies with many cool stars.  There are some
degeneracies between the fit parameters, e.g. the allocation of 
absorption EW between symmetric and blueshifted components;
the approach is only robust when the composite spectra are 
high S/N to constrain the symmetric component, which is
fit to one side of a Gaussian.  These issues do not affect our 
conclusions; we do find that both the
symmetric and the wind components have the same dependence
on galaxy properties -- they are stronger in more massive objects, 
as we show below.

The middle and upper panels of 
Figure \ref{fig-mgemsplit} show the coadded spectrum $F_{obs}$
of the 1356 galaxies excluding the excess emission sample, the 
symmetric \mgii\ doublet absorption (multiplied back by the
continuum), $C(1-A_{sym})$, and the outflow component
(also multiplied by $C$), $C (1-A_{flow}) F_{em}$.
Properties including EW and velocities of the components
are tabulated in Table \ref{table-absprops}.

\subsection{Optical thickness of the absorption}

In optically thin \mgii\ absorption, the 2796 \AA\ line
is twice as strong as the 2803 \AA\ line.  However, \mgii\
absorbers are frequently optically thick and saturated, in 
which case the line depths are close to
equal (e.g. Steidel \& Sargent 1992).

The symmetric Gaussian model goes deeper than the red wing
of the 2796 \AA\ absorption, which causes the inferred
outflow $F_{em} (1-A_{flow})$ (blue line in middle panel of 
Figure \ref{fig-mgemsplit})
to still show some excess emission, even though the strong
excess-emission galaxies have been weeded out.  The cause
must be excess emission, since the 2796 \AA\ symmetric line 
cannot be intrinsically weaker than the 2803 \AA\ line.

In the spectrum of Figure \ref{fig-mgemsplit} (middle panel),
the EWs of the outflow absorption component from 0 to -768 \kms\
are 1.54 and 1.37 \AA\ for the 2796 and 2803 \AA\ lines,
(excluding pixels above continuum to avoid diluting the 
blueshifted 2803 \AA\ line by emission in the 2796 \AA\ line).  
This yields a doublet ratio of 1.13,
implying quite saturated absorption.  Even this
is an upper limit since the 2803 blueshifted
absorption depth is still reduced by some emission in the 2796 line.
The absorption troughs reach equal depths at -100 \kms\ 
relative to each line, suggesting that the true line
ratio is close to 1.0 and the \mgii\ is highly saturated.
We discuss the depth and column density further in Section
\ref{sec-coldens}.
For saturated absorption, the equivalent width is more directly
controlled by the velocity spread than the column density.  

The \mgii\ absorption in both the symmetric and outflow
components is not optically thin, so
a distribution of column density $N_H(v)$ that is Gaussian
in velocity can not produce an optical depth or absorption
$A(v)$ that is Gaussian.  For a more detailed discussion of
the interaction of optical depth and covering fraction, see
Rupke \etal\ (2005a).  However, a superposition of a Gaussian
velocity distribution of small velocity-width clouds, each of 
which is optically thick but has a small covering fraction, can
produce a Gaussian line.  Stellar photospheric profiles can
also be fit by a Gaussian.

In practice, the coadded spectra
and blended doublet do not allow us to separate optical depth
and covering fraction.  A Gaussian is a reasonable fit to 
the red wing, while for the blueshifted absorption, the
coaddition of many spectra is analogous to an ensemble of 
absorbers over a range of optical depth and velocity spread,
as modeled by Jenkins (1986) in testing the applicability of
the classical doublet ratio method (Spitzer 1986).  
We use the ratio of the two \mgii\ lines to estimate the optical depth 
and column density in Section \ref{sec-coldens}.

As Figure \ref{fig-mgemsplit} shows, the \mgii\ absorption trough can
extend quite far to the blue, well beyond $-600$ \kms\ relative
to the 2796 \AA\ line.  It is difficult to measure the extent 
of blueshifted absorption in the 2803 \AA\ line because it
runs into the 2796 \AA\ line and excess emission.  In the
deepest part of the absorption trough, the two lines agree well.
If the lines become optically thin at high negative velocities,
the 2796 \AA\ line would be stronger, but this is difficult to
measure given blending and a possibly non-flat continuum.
The \mgi\ 2852 \AA\ line is weaker than the \mgii\ lines but 
has a similar velocity profile and ratio of blueshifted to symmetric
absorption.

\subsection{The sawtooth outflow profile}
\label{sec-windprofile}

The velocity profile of the \mgii\ blueshifted absorption
in the coadded spectrum shows a characteristic asymmetric
or ``sawtooth'' shape, with its deepest point around -150 \kms\
and lower absorption at large negative velocities out to nearly
-1000 \kms.  The \mgi\ profile has a similar shape.
This sawtooth shape also appears when the
sample is subdivided based on galaxy properties such as
mass and star formation rate (Section \ref{sec-propdepend}).
The asymmetric line shape is also seen in absorption in individual
galaxy spectra, as shown in Figure \ref{fig-samplespec}, 
Some individual galaxies could
have narrower, distinct absorption systems, as seen 
in high-resolution spectroscopy of low-redshift outflows
(Schwartz \& Martin 2004; Rupke \& Veilleux 2005;
see also the simulations of Fujita \etal\ 2008).  

In our coadded spectrum,
the sawtooth profile expresses the mean absorption in front
of star-forming galaxies as a function of velocity.
Individual absorbers are probably optically thick,
but the covering factor is less than 100\%,
so the mean absorption depth is essentially the probability
of finding a absorber at that velocity.
Because individual galaxies exhibit the asymmetric absorption
profile (at least in galaxies bright enough to see absorption
individually), it seems to be a property of the wind physics rather
than of the distribution of wind velocities over the ensemble.
This appears to rule out two toy models for the sawtooth
profile: (1) it is solely due to a distribution of 
characteristic wind velocities that has fewer winds at
high velocity; (2) the winds are well-collimated, and
the profile arises as we stack fewer galaxies with high outflow
velocities because face-on
galaxies are rarer.  We discuss an interpretation of the
sawtooth profile in terms of cool gas being accelerated 
by the wind in Section \ref{sec-windmodel}.

\section{Properties of the sample galaxies}
\label{sec-sampleprops}

The large number of galaxies in the Mg II sample allows
us to measure the dependence of outflows on host galaxy properties
by dividing the sample into several parts and coadding them
separately.   We compute restframe magnitude and color, and
stellar mass and star formation rate.  In this section, we 
describe how these are computed 
from the $BRI$ magnitudes and spectra, the calibration to
stellar masses from $K$ magnitudes, and calibration of UV SFRs 
to MIPS 24\um\ fluxes.  Because the Mg II sample is selected at 
restframe 2800 \AA, there are strong covariances among some
of the properties such as mass and star formation rate.
The distribution of the sample in several properties is shown 
in Figures \ref{fig-matrixplot} and \ref{fig-matrixplotsfr}.

\subsection{Magnitudes and colors}

To derive rest magnitudes, we use an SED interpolation procedure,
previously described and tested in Weiner \etal\ (2005) and Willmer 
\etal\ (2006).  For a DEEP2 galaxy at some redshift $z$,
we redshift the UV+optical SEDs of 34 actual galaxies from 
the atlas of Kinney \etal\ (1996), compute the desired restframe quantity,
e.g. restframe $U-B$ color, as a function of an observed color, e.g.
$R-I$, and combine the relation derived from the SEDs with the
actual $R-I$ color of the DEEP2 galaxy to infer its restframe
$U-B$ color.  At $z \sim 1.4$, the DEEP2 $BRI$ filters correspond
roughly to restframe 1800 \AA, 2800 \AA, and $U$-band.  Inferring
a blue magnitude $M_B$ and $U-B$ color effectively extrapolates
from a $2800-U$ observation, so the zeropoint of our $U-B$
color scale might suffer shifts of $0.1-0.2$ magnitudes when
compared to lower redshift data.  However, relative measurements
of color within our sample are accurate.  The UV magnitudes
$M_{2800}$ and $M_U$ are measured quite accurately by our
procedure since they correspond closely to the observed $RI$
filters.  We also estimate UV magnitudes at 1500 and 2200 \AA, 
close to the GALEX FUV and NUV bandpasses.

\subsection{Stellar masses}

Stellar masses are best derived from the infrared, but near-IR
photometry does not exist for our entire sample.  
For 33\% of galaxies with redshifts in the DEEP2 survey,
$K$ magnitudes and stellar masses were measured with the Palomar 
5-m telescope and WIRC camera, over $\sim 80$ nights of imaging
(Bundy \etal\ 2006; Conselice \etal\ 2008).  
The remaining two-thirds do not have $K$ magnitudes because the
Palomar/WIRC survey does not cover all of the DEEP2 area and
misses some fainter galaxies in regions where it is less deep.
We use the rest $M_B$ and $U-B$ to derive a stellar mass ($M_*$) 
estimate using color-$M/L$ relations (Bell \& de Jong 2001; 
Bell \etal\ 2003) and then empirically correct these to stellar 
masses computed using $K$-band photometry and SED fitting by
Bundy \etal\ (2006).  This is a revision of a similar calibration
computed in Lin \etal\ (2007).  
The details of the calibration and formulae which can be
used to derive $M_*$ for DEEP2 galaxies are given in 
Appendix \ref{app-mstellar}.

\begin{figure}[t]
\begin{center}
\includegraphics[width=3.5truein]{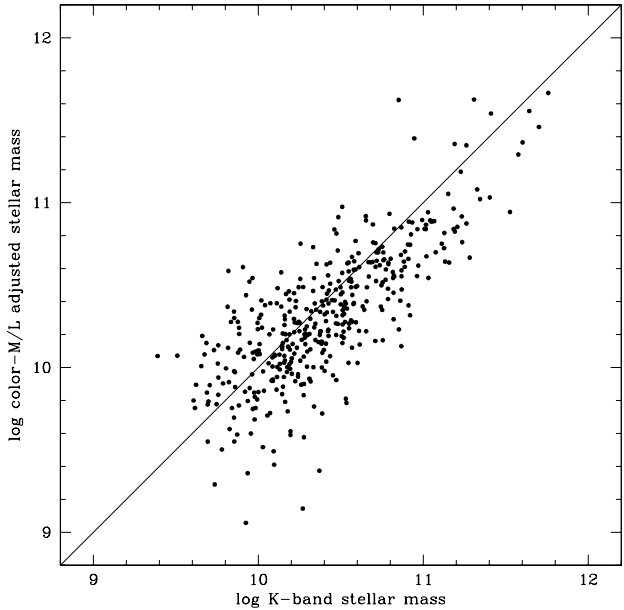}
\caption{
Stellar mass estimates from the SED fitting of Bundy \etal\ (2006)
including the $K$-band 
(x-axis) and from the optical color-$M/L$ relation empirically
corrected to $K$-band masses (y-axis).  There is an offset
of 0.09 dex and RMS
dispersion of 0.3 dex, but good overall agreement.
}
\label{fig-kmasscomp}
\end{center}
\end{figure}

\begin{figure*}[t]
\begin{center}
\includegraphics[width=5.5truein]{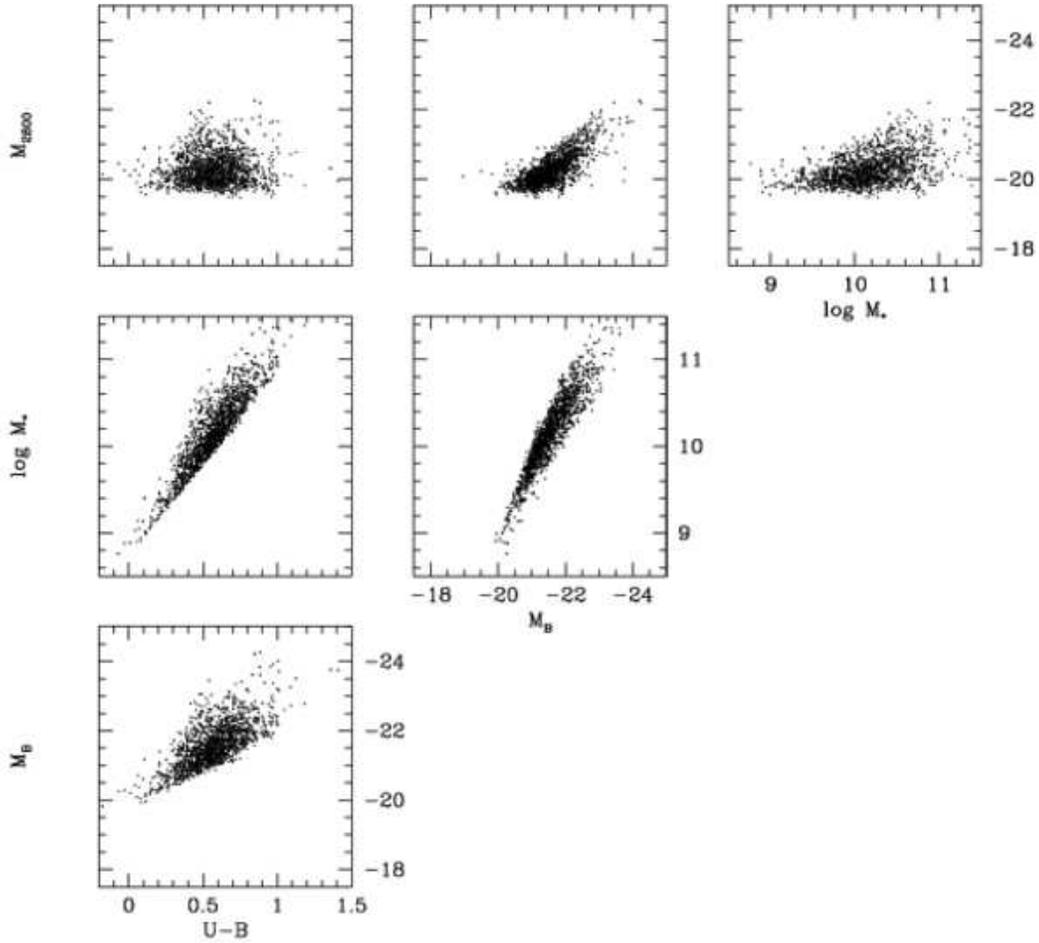}
\caption{Magnitude, color, and stellar mass estimates for galaxies 
in the Mg II sample.  The axes span 7.5 mag or 3 dex for magnitude
and mass.  The dependence of stellar mass on color means that the
sample spans a wider range of mass than UV or blue magnitude.}

\label{fig-matrixplot}
\end{center}
\end{figure*}

Figure \ref{fig-kmasscomp}
compares the resulting
mass estimates from color-$M/L$ to those from $K$-band;
both are for a Chabrier IMF.
The color-$M/L$ masses are 0.09 dex lower with an RMS
dispersion of 0.30 dex.  Although there may be 
remaining systematics, the relative order of $M_*$ is
reliable; our sample spans $\sim 1.5$ dex in stellar mass,
containing both low-mass blue
galaxies and higher-mass intermediate-color galaxies.  The
agreement in mass estimate allows using the color-$M/L$ relation
uniformly over the sample, since $K$ is not always available.
Figure \ref{fig-matrixplot} compares color, magnitude and stellar
mass for the sample.

\subsection{Emission line luminosities and velocities}

The primary emission line found in the Mg II sample is the \oii\
3727 \AA\ doublet.  We measure its intensity and width
as described briefly in Weiner \etal\ (2007):
in each DEEP2 1-d extracted spectrum,
we perform a non-linear least squares fit of a double Gaussian,
with the doublet line separation fixed but the intensity ratio
allowed to vary (although it is generally consistent with the
mean value of 1.4).  Each fit yields a measurement of the 
intensity of the doublet, its location, and the velocity
dispersion.  A similar procedure is used to measure
the velocity dispersion in the stacked spectra.

We compute a line equivalent width from the intensity
and a continuum measured over a 90 \AA\ window.
The EW and the calibrated $BRI$ photometry are combined to
obtain a total restframe line luminosity.  This procedure
bypasses instrumental throughput, and calibrates out slit
losses as long as the ratio of emission to continuum is 
the same inside the slit and outside.  For high-$z$ DEEP2 
galaxies, which nearly always have $r_{eff}<1\arcsec$,
the small size and seeing make this a safe assumption.

The typical \oii\ luminosity is high, $\sim 10^{42}$ erg s$^{-1}$,
and L(\oii) increases with $M_B$ or mass, as does the UV SFR estimate 
described below.  
However, we found that deriving star formation rates from \oii\
luminosity is quite dependent on assumptions about the metallicity and
extinction (e.g.\ Kewley \etal\ 2004; Weiner \etal\ 2007).  These are
unmeasured for this sample, and we expect 
evolution in them compared to lower-redshift samples (Shapley \etal\ 2005).  
Locally and at $z\sim 0.8$, there
are also strong trends in nebular line ratios as a function of mass
(Moustakas, Kennicutt \& Tremonti 2006; Weiner \etal\ 2007).
Without a calibration of these trends, SFR estimates from
\oii\ can be systematically off.  Applying a low-redshift
calibration (e.g.\ from Kennicutt 1998) produces SFRs that are
$\sim 0.5$ dex above the IR and far-UV SFR indicators discussed in the
next section.  We advise caution in applying \oii\ SFRs at
$z>1$ without better calibrations, which are beyond the scope
of this paper.

\subsection{Star formation rates}

Estimating star formation rates for high-$z$ galaxies 
generally relies on local calibrations and models.
No SFR estimators at high redshift are both sensitive and 
well tested; emission-line and UV estimators are affected
by extinction, while mid-IR surveys detect fewer objects
and may be affected by uncertainties in the IR SEDs.
We explore both IR and far-UV estimates of the SFR,
keeping in mind that there may be systematic effects.
In the end we are able to reconcile these estimates
by stacking the far-IR data.

\subsubsection{SFR from mid-infrared luminosity}

The mid- to far-IR luminosity emitted by dust heated by young stars
yields an estimate of the SFR that is not affected by extinction,
though it can be affected by AGN emission and by variations
in the IR SEDs.
Although only a small number of galaxies are detected in the IR,
the SFR(IR) is valuable both for calibrating UV and optical 
estimates of SFR, and for comparing to studies of outflows
in local IR-luminous galaxies (Heckman \etal\ 2000; Rupke \etal\ 2005a;
Martin 2005).

One DEEP2 field, the Extended Groth Strip, has been imaged
at 24 \um\ by scan mapping with MIPS, the Multiband Imaging
Photometer for Spitzer (Rieke \etal\ 2004), to a sensitivity
of 83 $\mu$Jy at $\sim 5\sigma$.  We matched the DEEP2 galaxies
to a catalog of 24 \um\ fluxes constructed by PSF fitting
(Davis \etal\ 2007).   Of the 1406-galaxy sample, 194 are
in the area imaged by MIPS at 24 \um\ and 38 galaxies are detected.

We estimate a total infared luminosity $\lir(8-1000 \um)$
from the 24 \um\ flux using an SED of Dale \& Helou (2002) 
with $\alpha=2.0$, which is applicable to star-forming galaxies
and follows the average trend of $8-24$ \um\ color with redshift
in DEEP2 galaxies.
Using this SED produces K-corrections similar to those
obtained by averaging a large number of Dale \& Helou SEDs,
as done by Bell \etal\ (2005).  At $z=1.4$, the conversion
is not a strong function of redshift, and 
$\rm{log}~\lir(\lsun) \simeq \rm{log}~S_{24} + 9.63$, with 
$S_{24}$ in \mujy.  The 83 \mujy\ limit corresponds to an
IR luminosity of $10^{11.5}~ \lsun$, and the median 
log \lir\ is 11.73.  34 of the 38 MIPS-detected 
sources are luminous infrared galaxies (LIRGs, $\lir >10^{11}~\sun$), and
four of the 38 galaxies have $\lir >10^{12}~\lsun$,
and are ultraluminous infrared galaxies (ULIRGs).

We estimate star formation rates by combining the UV(2800 \AA)
and IR luminosities, using the prescription of Bell \etal\ (2005):

\begin{equation}
SFR(IR) = 9.8 \times 10^{-11} (L_{IR} + 2.2 \nu L(2800\ {\rm \AA}))
\end{equation}

\noindent
where $L_{IR}$ and $L(2800) = 1.5\nu I_{\nu}(2800\ {\rm \AA})$ are 
in solar luminosities; this is defined for a Kroupa IMF (Kroupa 2001).
The IR term dominates this sum; in the MIPS-detected sample,
the median ${\rm log}~L_{IR}/(2.2\nu L(2800))$ is 0.9 dex.

Because the MIPS galaxies are IR-selected, they will have a 
larger $IR/UV$ than typical.  It is also possible that AGN
contribute to the IR luminosity in some of the objects.  However,
only one of the 38 MIPS sources is also among the 50
excess \mgii\ emission AGN objects.  
Another systematic effect is the use of a single typical star-forming 
IR SED to convert to total $L_{IR}$.  At $z=1.4$ the 24 \um\ 
band of MIPS is at restframe 10 \um, and the observed flux
could be enhanced by PAH emission or depressed by silicate
absorption.  Additionally, the dust temperature could be
different from local galaxies; Papovich \etal\ (2007) and
Rigby \etal\ (2008) find
that the 24 \um\ flux can overestimate the SFR in IR-luminous
$z \sim 2$ galaxies.  These issues are difficult to calibrate
out; they probably contribute of order 0.3 dex scatter to the
SFR(IR) (e.g. Bell \etal\ 2005), and there can be systematic 
overestimates of the SFR by a few tenths of a dex.  
In the next section we attempt to improve the SFR measurement
by comparing IR and far-UV, stacking the MIPS images to
get a median flux for non-MIPS-detected objects.

\subsubsection{SFR from UV luminosity}
\label{sec-uvsfr}

We can also estimate the SFR from UV luminosity.
At $z=1.4$, our $B$ and $R$ filters are roughly at restframe
1800 and 2800 \AA.  We compute absolute AB magnitudes at 1500 and 2200 \AA\
using our SED K-correction procedure, and compute the slope $\beta$ 
of the UV continuum, $f_\lambda \propto \lambda^\beta$.
The UV slope has been used as an indicator of extinction
by Meurer \etal\ (1999), who found 
a far-UV extinction at 1500 \AA: $A_{FUV} = 4.43 + 1.99\beta$
for local starburst galaxies.  This relation may vary depending
on galaxy type; Bell (2002) found that local normal galaxies deviate
to lower $A_{FUV}$.  Large samples of local galaxies studied in 
the UV with GALEX suggest that applying the Meurer 
\etal\ (1999) prescription can overestimate $A_{FUV}$ 
(Seibert \etal\ 2005; Treyer \etal\ 2007; Salim \etal\ 2007).

Although our $z=1.4$ sample is selected at 2800 \AA\ and is 
forming stars rapidly, it is not clear that local starbursts are
the best approximation as these are a highly selected sample.
We computed FUV attenuation using the relation
$A_{FUV} = 3.16 + 1.69 \beta$ (Seibert \etal\ 2005, as updated
by Treyer \etal\ 2007).  We correct the FUV luminosity
and compute an SFR following Kennicutt (1998), scaled by
0.7 to convert from a Salpeter to Kroupa IMF for consistency
with the SFR(IR) estimate:

\begin{equation}
SFR(FUV) (\msunyr) = 1.0 \times 10^{-28} L_\nu(1500\ {\rm \AA}) (\rm{erg/s/Hz}).
\end{equation}

The $A_{FUV}$ prescription of Salim \etal\ (2007) 
yields nearly identical SFR,
while the prescription of Meurer \etal\ (1999) yields SFRs
0.4 dex higher.  The accuracy is limited since there is
substantial scatter about the $A_{FUV}-\beta$ relation, 
0.9 mag in the Seibert \etal\ (2005) sample. 
The median values and $\pm 34\%$ range for $\beta$ and
$A_{FUV}$ in our no-excess-emission galaxy sample are $-0.65 \pm 0.42$
and $2.06 \pm 0.71$ respectively.  The UV continuum slope
is correlated with magnitude and restframe color - brighter,
redder, more massive galaxies have redder UV continua.

\begin{figure}[t]
\begin{center}
\includegraphics[width=3.5truein]{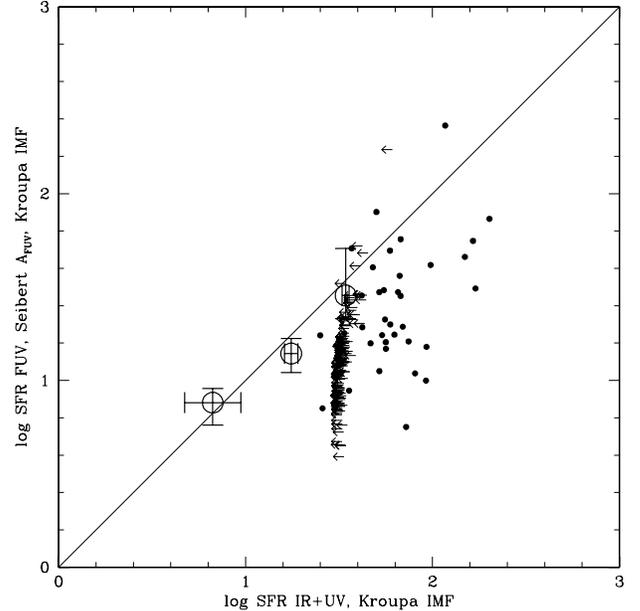}
\caption{
Comparison of star formation rates inferred from 24 \um\ + UV,
and from far-UV luminosity with the Seibert \etal\ (2005) relation
for $A_{FUV}$, for 194 galaxies in 
the \mgii\ sample with MIPS imaging. Calibrations refer to a Kroupa IMF.
Leftward-pointing arrows are upper limits for galaxies
not detected by MIPS.  The large open circles show the
median SFR(FUV) for galaxies subdivided into three ranges of $L_{FUV}$,
and the SFR(IR+UV) derived by a median stacking of their MIPS 24 \um\ images.
}
\label{fig-sfr_uvircomp}
\end{center}
\end{figure}

Figure \ref{fig-sfr_uvircomp} shows a comparison of SFR 
estimates from the far-IR and from the far-UV with the 
Seibert \etal\ (2005) $\beta-A_{FUV}$ relation, for 194
galaxies in the \mgii\ sample and MIPS imaging area.
For galaxies not detected by MIPS, we set an upper limit
based on the $4\sigma$ limit at 24 \um\ of 65 \mujy.

Because many of the galaxies are not individually detected
by MIPS, we divided the FUV sample into three ranges 
at the 25 and 75th percentiles.  For those galaxies with MIPS
imaging, we then median stacked the MIPS 24 \um\ images.
This yields a good detection in all three ranges.  The 
median SFR(IR+UV) for the three ranges is plotted against the 
median SFR(FUV) as the large circles in Figure \ref{fig-sfr_uvircomp}.

\begin{figure*}[t]
\begin{center}
\includegraphics[width=5.5truein]{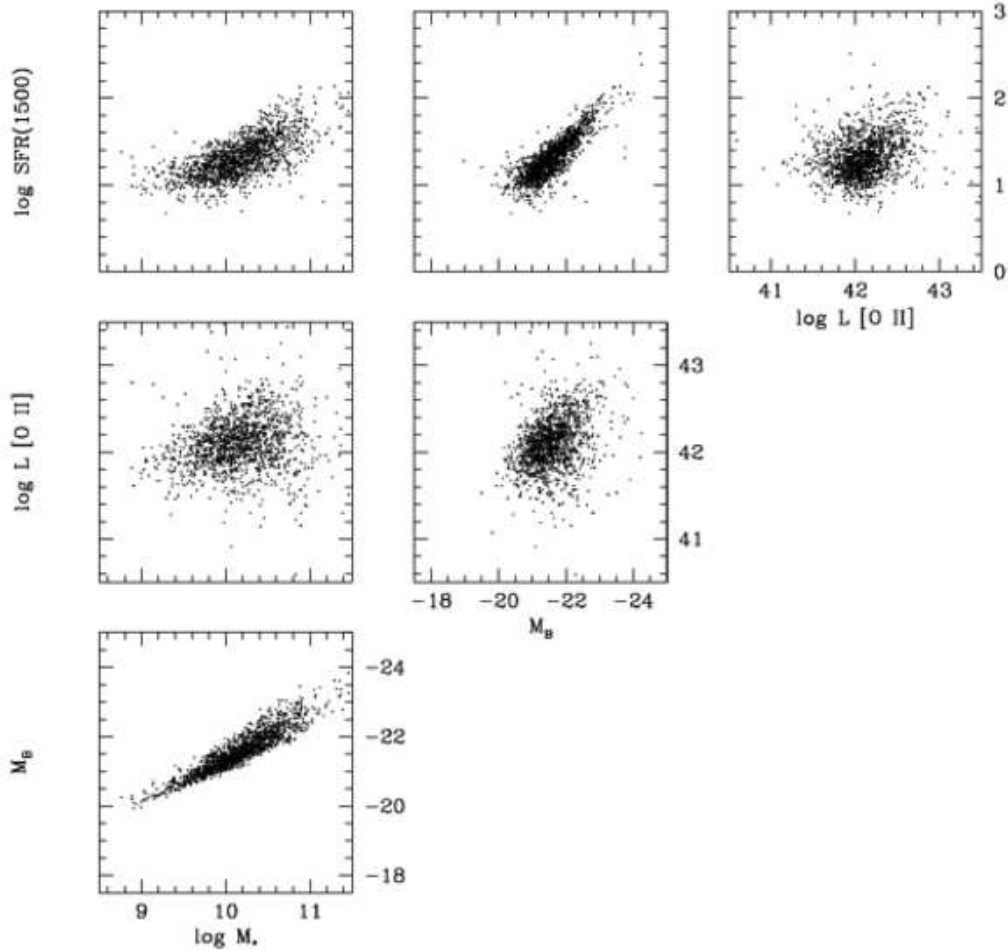}
\caption{Blue magnitude and stellar mass compared to \oii\ luminosity
in erg/sec, and star formation rate estimated from the FUV luminosity
in \msunyr.  The SFR(FUV) is corrected for extinction and
the \oii\ luminosity is not.  The extinction correction causes $M_B$
and $SFR(FUV)$ to be highly correlated.}
\label{fig-matrixplotsfr}
\end{center}
\end{figure*}

The SFR(FUV) following Seibert \etal\ (2005) is in  
good agreement with the SFR(IR); it is 
0.1 dex low for the more rapidly star-forming galaxies.
The Meurer \etal\ (1999) prescription for $A_{FUV}$ yields SFRs 0.4 dex
higher than the Seibert $A_{FUV}$, and would overestimate SFR
by $\sim 0.3$ dex for this sample.  
We adopt the SFRs derived from the Seibert $A_{FUV}$ and plot them in
Figure \ref{fig-matrixplotsfr}.
These galaxies are on average higher SFR, more
IR-luminous, and can be more extincted than the local
samples for which the UV relations were defined.
However, the stacked-IR to UV comparison allows
us to infer an SFR for all galaxies in the sample that is 
consistent with mid-IR determinations, even though only
a small number of the galaxies have individual mid-IR measurements.
As with the mid-IR measurement, AGN luminosity could contribute
to the far-UV, but only a small fraction of the galaxies have
evidence for an AGN, based on \mgii\ emission.  In addition,
of the sample galaxies in the EGS field,
very few are detected in X-rays (Nandra \etal\ 2007).

The stacked MIPS detections imply that the galaxies in the
$<25$\%ile, $25-75\%$ile, and $>75\%$ile of SFR(FUV) have 
median IR luminosities 
$\lir(8-1000\mu m) = 3.1 \times 10^{10}, 1.3 \times 10^{11},$
and $2.5 \times 10^{11}~\lsun$ respectively, and median
SFRs of 7, 18, and 34 \msunyr.  The median galaxy in the
$z\sim 1.4$ \mgii\ sample is roughly at the LIRG threshold of
$\lir = 10^{11}~\lsun$.

\subsection{Correlation of galaxy properties with SFR}

Figure \ref{fig-matrixplot} plots galaxy 
magnitudes, stellar mass, and color
against each other, and Figure \ref{fig-matrixplotsfr} compares
magnitude and stellar mass estimates against star formation indicators.
Because the sample is selected at restframe
2800 \AA, color, $M_B$ and stellar mass are highly covariant.
The color-magnitude plot shows that the sample contains blue
galaxies at a range of luminosities, but few red-sequence
galaxies; red galaxies are disfavored by the UV selection and
their lower abundance at $z>1$ (Willmer \etal\ 2006).
SFR as estimated from the UV is strongly
correlated with $M_B$ but somewhat less well with stellar mass:
galaxies with high extinction-corrected FUV luminosity are fairly 
high-mass.  The SFRs are for a Kroupa IMF while the
stellar masses are normalized to a Chabrier IMF, which is
just slightly lower.

Note that the $SFR_{FUV}$ correlates fairly tightly with $M_B$,
more so than with other quantities; $SFR_{FUV}$ is less tightly
correlated with UV luminosity itself.  This is caused by the 
extinction correction - the UV slope and $UV-B$ color are highly
correlated (see also Treyer \etal\ 2007).  The correlation is
also partially induced by the K-correction procedure since
we have to infer $M_B$ from restframe $U$ and color.  If the sample 
included red non-starforming galaxies, they would fall to lower SFR,
below the correlation.

The velocity linewidth $\sigma$ of the \oii\ emission is correlated
with magnitude and stellar mass, with considerable scatter.
Much of this scatter is due to the large scatter in the
high-redshift linewidth Tully-Fisher relation (Weiner \etal\ 2006),
which may be induced both by inclination and by details of
the internal kinematics.  
Velocity width and SFR are essentially uncorrelated, largely
due to scatter in the estimators.


\section{Outflow dependence on galaxy properties}
\label{sec-propdepend}

To study how galaxy properties affect the outflow, we 
divide the no-excess-emission sample into subsamples binned
in either magnitude, stellar mass, or star formation rate
in turn, and coadd the spectra with these bins.
We choose the bin limits to separate the bottom 25\%, middle
25\%-75\%, and top 25\% in each property.

\subsection{Outflows exist across the range of luminosity, mass, and SFR}

\begin{figure*}[t]
\begin{center}
\includegraphics[width=7.0truein]{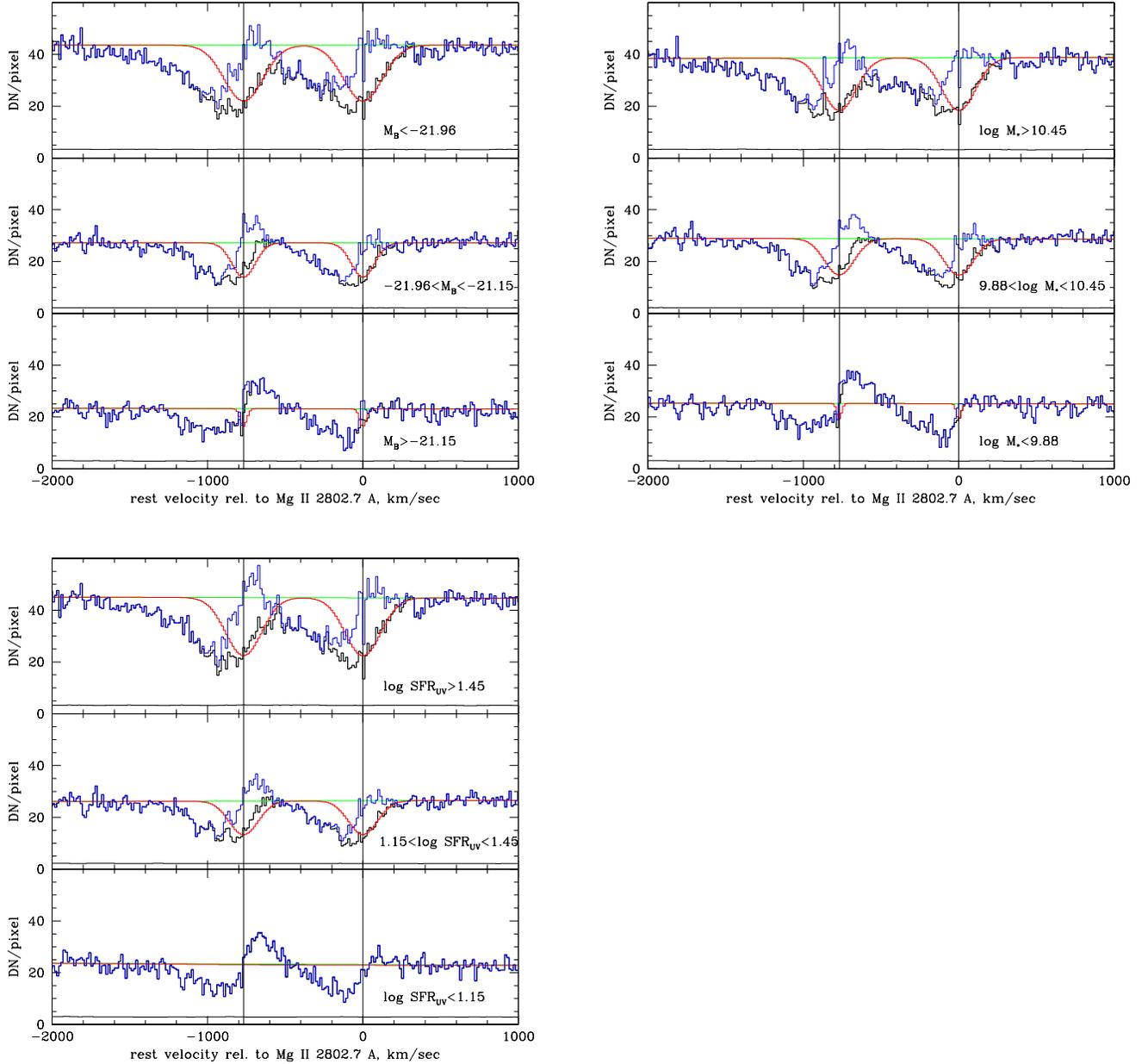}
\caption{
\mgii\ absorption for subsamples split by galaxy properties.
In each panel, co-added spectra are shown for three subsamples, of low
quartile, middle 50\%, and high quartile.  The subsamples
are divided on: $M_B$ absolute magnitude (upper left panel),
$M_*$ stellar mass (upper right), and $SFR_{FUV}$ (lower left).
The co-added spectra are decomposed into continuum (green straight line), 
symmetric absorption (red gaussians), and blueshifted absorption (blue 
asymmetric line),
as in Figure \ref{fig-mgemsplit}.
}
\label{fig-3subsplit}
\end{center}
\end{figure*}

The panels of Figure \ref{fig-3subsplit}
show the coadded spectra at \mgii\
for subdivisions in $M_B$, $M_*$, and $SFR(FUV)$ respectively.
Table \ref{table-absprops} summarizes the results of the 
decomposition into symmetric and outflow absorption, and
the \oii\ emission in each coadded spectrum.

The symmetric absorption is strongly dependent on the
properties, being much weaker in the low-mass, low-SFR
end of the sample.  The dispersion of the 
symmetric \mgii\ absorption varies more strongly than the
velocity dispersion of \oii; it is high in redder/more massive
objects.  These objects contain a cooler stellar population,
with stronger and broader photospheric \mgii\ absorption
(as in e.g. the models of Rodr{\'{\i}}guez-Merino \etal\ 2005),
so that the dispersion reflects stellar properties.
Note that there is almost no absorption at zero velocity
in the bluer, lower-mass, lower-SFR subsample.  This is
surprising since internal absorption
by the galaxies' ISM should occur, on average, at zero velocity,
and even low columns of cool gas produce \mgii.
It appears that in the lower-mass galaxies {\it all} of the 
cool gas between us and the UV-emitting stars is outflowing.

The outflow component of the absorption exists in all subdivisions
of the sample.  Its equivalent width is higher in the brighter,
redder, high-mass, high-SFR subdivisions -- since all of these 
galaxy properties are correlated, it is not possible to isolate
one parameter that predicts outflow strength much better than the
others.  Although the blueshifted absorption is stronger in the
larger galaxies, it is still fairly strong in the smaller, low-mass
galaxies.  We have not found any class of galaxy in the sample 
that does not exhibit outflow absorption, justifying the claim
above that the \mgii\ outflows are ubiquitous in the DEEP2 sample.
Although this sample is selected to all be fairly luminous at
restframe 2800 \AA, it covers a range of $10\times$ in SFR and
$B$ luminosity, and $30\times$ in stellar mass.

We quantify the velocity extent of the outflow absorption in two ways:
the median velocity of absorption, $V_{med}$, and the velocity
where outflow-component absorption reaches a threshold depth.  We
lightly smooth the outflow absorption spectrum $A_{flow}$
with a boxcar of 5 pixels (68 \kms) to reduce noise, and
count from $-100 \kms$  downward to more negative velocities
until reaching a pixel that is above 0.75
or 0.9 of the continuum value.  These velocities define the
locations $V_{25\%}$ and $V_{10\%}$, respectively, the extent
of contiguous 25\% or 10\% absorption depth.  
These threshold measurements
are more robust than attempting to measure a maximum outflow 
velocity, since in the composite spectra, the outflow absorption 
is not a set of discrete components, but gradually asymptotes to zero at
high velocities. 

Figure \ref{fig-windcomp} overplots the absorption components
of the three subsamples in mass and SFR, demonstrating that
the high-mass and high-SFR subsamples show greater 
absorption at $-800 < V < -400$ \kms\ than the intermediate
and low-mass or -SFR subsamples. 
The velocity extents of outflow are tabulated in 
Table \ref{table-absprops} and plotted in Figures \ref{fig-velmass}
and \ref{fig-velsfr}.

Both the depth of the outflow absorption and its
velocity extent are greater for the brighter, high-mass, 
high-SFR, redder-color fractions of the sample.
This is similar to the trend found by Shapley \etal\ (2003)
in $z\sim 3$ Lyman-break galaxies, where low-ionization lines
are stronger in the galaxies with higher SFR, larger velocity
offset, and redder UV continuum slope.  Fitting a 
power-law dependence of the velocity of the 10\% absorption depth, 
$V_{10\%}$, on mass or SFR, yields $V_{10\%} \propto M_*^{0.17}$
and  $V_{10\%} \propto SFR_{FUV}^{0.38}$.  The power-law slopes for 
$V_{25\%}$ are 0.11 and 0.17 on $M_*$ and $SFR_{FUV}$ respectively.
These measures are dependent on the choice of the absorption
threshold, but it appears that that $V_{wind} \sim SFR_{FUV}^{0.3}$.
It is suggestive that the velocity appears to have a stronger
dependence on SFR than on stellar mass, but this could be
because our sample spans a larger range in stellar mass than
it does in SFR.  Notably, the $z=1.4$ starforming galaxies have a
similar dependence of wind velocity on SFR to the low-z ULIRGs studied
by Martin (2005) in Na I, who found an upper envelope
of wind velocity at $V \sim SFR^{0.35}$.

\subsection{Search for outflow dependence on galaxy type}
\label{sec-typedepend}

Locally, starburst-driven winds are more frequently seen in face-on
galaxies, due to collimation of biconical outflows 
(Heckman \etal\ 2000; Rupke \etal\ 2005b).  
118 of the \mgii\ sample galaxies are in the Extended Groth Strip and have 
HST ACS $I$-band imaging (Davis \etal\ 2007; Lotz \etal\ 2008). 
We divided them up by axis ratio as measured with SExtractor 
(Bertin \& Arnouts 1996)
in an attempt to separate face-on and edge-on galaxies.  
We did not find a correlation between axis ratio and wind strength
or wind velocity. 
An examination of the ACS images
shows that the galaxies are too irregular and low surface brightness
to trust the axis ratio as a measure of disk inclination.  Many may
not be disks.  Because the imaging is at restframe $U$-band,
the high-z galaxies are more irregular, and their images appear 
to be highly affected by starforming regions and by dust.

We also classified the galaxies by eye into six broad
morphological types:
face-on, edge-on, chain, irregular, merger, and compact, 
finding 37, 13, 18, 28, 3, and 19 respectively.  The large
number of chain, irregular and compact galaxies indicates that
many $z=1.4$ galaxies are not obviously disky in the rest $U$.
When we coadd the spectra within each type, we find that
all types exhibit blueshifted \mgii\ absorption -- even the 
3 mergers have enough signal to detect it.  However, the noise
in the spectra and the variance among individual objects
within a type prevent us from determining whether the outflows
in one type are stronger than another.  Note however that 
the merger fraction is low, as is true of $z\sim 1$ galaxies
in general (Lotz \etal\ 2008).  While many galaxies are irregular
or disturbed, major mergers are not a prerequisite for
driving a wind in this sample.

In the ACS imaging, the median of the
galaxies' Petrosian radii is 5.2 kpc and there is a weak
correlation between size and magnitude, and no obvious
correlation between size and outflow.
These tests show that 
winds are found across all star-forming types at $z=1.4$, but 
they neither prove nor disprove collimation of winds
in this sample.  A larger number of galaxies with HST near-IR imaging, 
and spectra at \mgii, are needed
to search for collimation and morphological type dependence.

\begin{deluxetable*}{rrrrrrrrrrrrrrrr}
\tablefontsize{\scriptsize}
\tablecaption{
Absorption and emission properties of co-added galaxy samples
\label{table-absprops}
}

\tablecolumns{16}
\tablewidth{0pt}
\tabletypesize{\scriptsize}

\setlength{\tabcolsep}{0.0in}

\tablehead{
 category  & \# of & \multicolumn{2}{c}{Symmetric absorption\tablenotemark{a}} & 
   \multicolumn{4}{c}{Outflow absorption\tablenotemark{a}} & 
   \multicolumn{2}{c}{\oii\ emission\tablenotemark{c}}
 \\
    & galaxies  & $W_{2803}$, \AA & $\sigma$\tablenotemark{b}, \kms &
    $W_{2796}$, \AA & $V_{med}$ & $V_{25\%}$ & $V_{10\%}$ &
    EW\tablenotemark{d}, \AA & $\sigma$\tablenotemark{b}
 }
 \startdata
All, \mgii\        & 1406 & $0.61 \pm 0.03$ & $64.7 \pm 4.1$  &
 \quad$1.71 \pm 0.06$ & $-279^{+17}_{-9}$  & $-364$ & $-515$  &
 $44.3$ & $68.7$ \\
All, \mgi\         & 1406 & $0.32 \pm 0.04$ & $106.0 \pm 15.9$  &
 \quad$0.58 \pm 0.05$ & $-194^{+33}_{-19}$ & ... & $-299$  &
 $44.3$ & $68.7$ \\
\tableline
MIPS 24 \um\ sources &  37 & $0.64 \pm 0.12$ & $69.6 \pm 16.2$ &
 \quad$1.70 \pm 0.22$ & $-311^{+59}_{-37}$ & $-501$ & $-584$  &
 $30.2$ & $67.6$ \\
\tableline
$M_B>-21.15$ & 339 & $0.14 \pm 0.05$ & $0.0$ &
 \quad$1.49 \pm 0.16$ & $-266^{+36}_{-49}$ & $-419$ & $-474$ &
 $64.2$ & $59.1$ \\
$-21.15<M_B<-21.96$ & 678 & $0.85 \pm 0.05$ & $70.2 \pm 5.4$ &
 \quad$1.45 \pm 0.10$ & $-226^{+23}_{-24}$ & $-336$ & $-405$ &
 $47.9$ & $69.6$ \\
$M_B<-21.96$ & 339 & $1.44 \pm 0.07$ & $122.0 \pm 6.7$ &
 $1.96 \pm 0.10$ & $-315^{+33}_{-10}$ & $-405$ & $-639$ &
 $31.1$ & $76.2$ \\
\tableline
${\rm log} M_*<9.88$ & 339  & $0.08 \pm 0.04$ & $0.0$ &
 \quad$1.39 \pm 0.14$ & $-269^{+43}_{-37}$ & $-295$ & $-446$ &
 $63.2$ & $60.3$ \\
$9.88<{\rm log} M_*<10.45$ & 678 & $1.08 \pm 0.06$ & $91.3 \pm 5.9$ &
 \quad$1.55 \pm 0.10$ & $-246^{+34}_{-8}$ & $-336$ & $-515$ &
 $50.2$ & $69.6$ \\
${\rm log} M_*>10.45$ & 339 & $1.33 \pm 0.07$ & $104.7 \pm 7.6$ &
 \quad$1.91 \pm 0.12$ & $-309^{+23}_{-27}$ & $-377$ & $-653$ &
 $27.6$ & $76.7$ \\
\tableline
${\rm log} SFR <1.15$ & 339 & $-0.13 \pm 0.30$ & $0.0$ &
 \quad$1.46 \pm 0.15$ & $-233^{+43}_{-32}$ & $-295$ & $-432$ &
 $60.1$ & $60.2$ \\
$1.15<{\rm log} SFR <1.45$ & 678 & $1.08 \pm 0.06$ & $90.2 \pm 6.4$ &
 \quad$1.43 \pm 0.11$ & $-256^{+30}_{-21}$ & $-336$ & $-446$ &
 $47.2$ & $69.0$ \\
${\rm log} SFR >1.45$ & 339 & $1.34 \pm 0.06$ & $111.8 \pm 6.2$ &
 \quad$1.84 \pm 0.10$ & $-290^{+10}_{-32}$ & $-364$ & $-653$ &
 $33.3$ & $75.0$ \\

 \enddata
\tablenotetext{a}{Symmetric and outflow absorption are measured for
\mgii\ 2803 and 2796 \AA, respectively, except for the ``All, \mgi''
sample, where they are for \mgi\ 2852 \AA.  Absorption EWs are for 
the  2796 \AA\ line of the \mgii\ doublet; the total is approximately
twice as large.}
\tablenotetext{b}{Corrected for $\sigma_{inst} =25 \kms$.  
Symmetric absorption lines in the faint/blue quartile are consistent 
with zero intrinsic width.}
\tablenotetext{c}{Statistical errors on \oii\ emission are $0.5-1$ \AA\ in EW
and $0.5-2$ \kms\ in $\sigma$.}
\tablenotetext{d}{\oii\ emission EW is the total of both lines in the
doublet.}


\end{deluxetable*}

\section{Analysis and Discussion}
\label{sec-discussion}

\subsection{Physical properties of the outflow}
\label{sec-coldens}

The amount of gas contained in the outflows is of
great interest in gauging its importance to galactic and
IGM evolution.  Because the \mgii\ lines are optically thick,
approaching the ``flat part'' of the curve of growth,
we cannot infer column density directly from the equivalent
width.  Here we use the classical doublet ratio method
to approximate the optical depth and column density
(Spitzer 1968; see also Jenkins 1986 and Rupke \etal\ 2005a).

The EW ratio of the two \mgii\ lines, $W_{2796}/W_{2803}$,
varies from 2 to 1 for optical depth $\tau$ from 0 to infinity. 
For a single absorber, there is a relation between
the EW ratio and the optical depth at line center $\tau_0$
($C$ in the notation of Spitzer 1968, Table 2.1).
For an ensemble of absorbers, the relation does not strictly
apply, but generally yields accurate results in the mean if the
saturation is not too strong (Jenkins 1986).
The doublet ratio is given by $F(2\tau_0)/F(\tau_0)$ where
$F$ is the integral of transmission over the line, and the
column density in atoms \cmsq\ is (Spitzer 1968):

\begin{equation}
{\rm log}~N = {\rm log} \frac{W_{2803}}{\lambda} - 
{\rm log} \frac{2F(\tau_0)}{\pi^{1/2}\tau_0} - {\rm log} \lambda f_{2803} + 20.053.
\end{equation}

We computed EWs for the \mgii\ lines in the outflow components
of the no-excess-emission sample (1356 galaxies),
and in the subsamples split by stellar mass and SFR.
We compute the EW between 0 and -768 \kms\ for each line,
and exclude above-continuum pixels to eliminate those
most grossly affected by emission.  
These EWs and the inferred optical depths 
are tabulated in Table \ref{table-ewratio}.  
The formal errors on the EWs are small, but the
systematics are significant due to (1) underlying \mgii\ emission,
which tends to decrease $W_{2803}$ and overestimate the
true ratio; and (2) limits on the applicability of the simple doublet 
ratio method.  Given systematics, there are no significant
differences between the subsamples in doublet ratio or in
optical depth.

Generally, the doublet ratios derived 
are 1.1 to 1.2, or even lower.  Adjusting the velocity
interval has little effect.  Computing the ratio
directly on the observed spectrum, rather than the wind
component only, yields lower doublet ratios, around 1.0.

\begin{deluxetable}{rrrll}
\tablecaption{
Mg II equivalent width and doublet ratios 
\label{table-ewratio}
}

\tablecolumns{5}
\tablewidth{0pt}
\tablehead{
 category  & $W_{2796}$\tablenotemark{a}, \AA & $W_{2803}$\tablenotemark{a}, \AA & Ratio & $\tau_0$ 
}
\startdata

no excess emission         &  1.54  &  1.37  &  1.13  & $\sim 8-10$  \\
${\rm log} M_*<9.88$       &  1.24  &  1.26  &  0.98\tablenotemark{b} & $>10$ \\
$9.88<{\rm log} M_*<10.45$ &  1.55  &  1.41  &  1.10  & $\sim10-20$ \\
${\rm log} M_*>10.45$      &  1.73  &  1.48  &  1.17  & $\sim 5$ \\
${\rm log} SFR <1.15$      &  1.49  &  1.25  &  1.19\tablenotemark{b} & $\sim 5$ \\
$1.15<{\rm log} SFR <1.45$ &  1.37  &  1.31  &  1.04  & $>10$ \\
${\rm log} SFR >1.45$      &  1.67  &  1.47  &  1.14  & $\sim 8$ \\

\enddata
\tablenotetext{a}{EW is computed only between 0 and -768 \kms\ and
excludes above-continuum pixels.}
\tablenotetext{b}{Most strongly affected by emission.}
\end{deluxetable}

For doublet ratios $1.20-1.10$, the optical depths are $\tau_0 = 4-20$,
and $F(\tau_0)/\tau_0 = 0.33-0.093$, so there is a factor of $\sim 3.5$
range in inferred column density. 
To get a characteristic column density of the outflow, we take typical
values $W_{2803} = 1.37$~\AA\ and $\tau_0= 10$, with 
${\rm log} \lambda f$ = 2.933 (Morton 1991).  The
inferred column density is ${\rm log}~ N_{Mg II} = 14.5$ atoms \cmsq.
The optical depths derived here, $\tau_0 \sim 4-20$, 
range high enough that the results of Jenkins (1986) may not
apply and the inferred column densities could be 
{\it underestimated.}

To be conservative, we make no ionization correction; \mgi\ is present but
relatively weak.  The presence of \mgi\ in the outflow, with ionization
potential 7.7 eV, suggests some amount of cold gas and/or
shielding from UV radiation.  The \mgi\ line has a similar profile 
to \mgii, but the \mgi\ is not as deep.  \mgi\ could be weaker 
due to lower column density or lower covering fraction.
Because \mgii\ is saturated
and \mgi\ has a higher oscillator strength than \mgii, getting the
observed \mgi\ by lowering the column requires fine-tuning the
$\mgi/\mgii$ ratio to a precise value to obtain unsaturated
absorption.  A more likely explanation is that \mgi\ is 
saturated but has lower covering fraction than \mgii; similarly
we found a higher covering fraction for \mgii\ than Rupke \etal\
(2005b) find for Na I in low-redshift outflows.
A modeling of the \mgii, \mgi, and
incident UV field is beyond the scope of this paper, but could
yield interesting constraints on wind physics (cf. Murray \etal\ 2007).

Assuming a solar abundance ratio of 
${\rm log~Mg/H} = -4.42$ and an Mg depletion of --1.2 dex,
typical in the local ISM (Savage \& Sembach 1996) and reasonable
for outflows given that local outflows in luminous galaxies
are dusty (Heckman 2002), the
inferred atomic column density is $N_H = 1.3 \times 10^{20}$ atoms \cmsq.
The average galaxy in the DEEP2 $z \sim 1.4$ sample 
is expelling and covered by at least a sub-damped Lyman-alpha absorber's 
worth of outflowing gas.  This column density is integrated over 
viewing angle and clumpiness of absorbers, so it includes the covering
fractions $C_\Omega$ and $C_f$.
The column density is moderately
lower than the columns of $\sim 10^{21}$ atoms \cmsq\
inferred from Na I in low-redshift
infrared-luminous galaxies (Heckman \etal\ 2000;
Rupke \etal\ 2005b).  Our estimate is conservative in the
optical depth and lack of ionization correction, and possible
effects of highly saturated \mgii, while those
authors have had to apply a large ionization correction for Na I,
so the measurements are essentially consistent.  

The IR luminosities
of the Heckman sample are roughly comparable to the DEEP2 galaxies, 
while the Rupke sample is at higher $L_{IR}$ in the mean,
since only half of the DEEP2 galaxies would be LIRGs.  The major
difference is frequency of occurrence: the Heckman and Rupke samples
are relatively special objects locally, while the DEEP2 sample 
represents a broad range of the luminous star-forming objects 
at $z \sim 1.4$, where the global SFR is higher.  This reflects the
overall higher SFR of individual galaxies at high redshift
(e.g. Le Floc'h \etal\ 2005; Noeske \etal\ 2007a), but there may also
be differences in the physical properties of the objects,
compared further in Section \ref{sec-localcomp}.

The column density, velocity and a characteristic size allow
an estimate of the typical mass outflow rate.  For a 
single thin shell wind at radius $R$, thickness $d<<R$, 
outflow velocity $v$, with angular covering fraction $C_\Omega$,
clumpiness covering fraction $C_f$,
and mean atomic weight $\mu m_p$,

\begin{equation}
\dot M = 4 \pi C_\Omega C_f \mu m_p N_H R v,
\end{equation}
\noindent
while for a thick wind that extends from radius 0 to $R$,

\begin{equation}
\dot M = \frac{4 \pi}{3} C_\Omega C_f \mu m_p N_H R v.
\end{equation}

For a composite spectrum of galaxies viewed at all angles, the
composite integrates over covered and uncovered lines of sight and
clumpiness within the outflow,
so we set $C_\Omega C_f=1$.
The sizes of the outflows are not easily constrained but
must be at least of order the galaxy radius, since the
covering factor is large.  In our sample, the median
Petrosian radius for galaxies with ACS imaging is 5.2 kpc
and the 50\% outflow velocity is 279 \kms.  For a thin shell
geometry and using typical numbers, with $\mu=1.4$:

\begin{equation}
\dot M \simeq 22~\msunyr  \frac{N_H}{10^{20}~\cmsq} \frac{R}{5~{\rm kpc}} \frac{v}{300~\kms}.
\end{equation}

This mass outflow rate is similar to the galaxies' star formation rates
of 10--100 \msunyr, supporting the idea that galactic winds powered
by star formation and the resultant supernovae
drive an outflow of $\dot M \sim SFR$ (e.g. Heckman \etal\ 2000; 
Pettini \etal\ 2000; Heckman 2002; Veilleux \etal\ 2005).  
A mass outflow rate roughly equal to the SFR, fueled by continuing
accretion, can help explain trends of SFR and metallicity in
high-redshift galaxies (Erb 2008).
The star formation rate, characteristic velocity, and outflow EW are 
all higher in the more massive/luminous objects in our sample, so
the mass outflow in large galaxies is substantial.

A critical unknown is the radial extent that the outflow reaches.
Some dramatic instances of superwinds in local galaxies that reach large 
distances are seen in X-ray and H$\alpha$ imaging 
(e.g. Strickland \etal\ 2004).
It is well known that \mgii\ QSO absorption line systems occur
within $\sim 50$ kpc of luminous galaxies, and also
that luminous galaxies have a large covering factor within
that radius, possibly dependent on galaxy type
(e.g. Lanzetta \& Bowen 1990; 
Bergeron \& Boisse 1991; Steidel, Dickinson \& Persson 1994; 
Bowen, Blades \& Pettini 1995; Chen \& Tinker 2008).
However, whether these systems are mostly related to galactic
outflows is debated (e.g. Bond \etal\ 2001; Steidel \etal\ 2002;
Prochter, Prochaska, \& Burles 2006;
Bouch{\'e} \etal\ 2006; Kaprczak \etal\ 2007; Tinker \& Chen 2008).  
From these examples, it is 
suggestive, but not definitive, that the \mgii\ outflows
could reach large distances, tens of kpc or more.  In the next section
we use outflow velocities to discuss whether the gas can escape.

\subsection{Can winds escape from galaxies?}
\label{sec-escape}

Galactic winds are the prime suspect for the enrichment
of the intergalactic medium.  The widespread detection
of metals in intergalactic and intracluster gas reveals
that, somehow, gas must get out of galaxies.  The
details of this process are difficult to model, and a
number of treatments have come to different answers on
the significance of gas and metal escape as a function
of galaxy mass.  Some have favored gas escape from the
shallower potential wells of small galaxies and dark halos
(Larson 1974; Dekel \& Silk 1986) while others suggest
that the ISM dynamics and loading of cool gas into the
wind are more important than the depth of the potential
(De Young \& Heckman 1994; Strickland \& Stevens 2000).

Here the best probe we have is to compare the wind 
velocity to the galaxy escape velocity; we assume
that the observed high velocity wind gas has already punched
out of the disk or densest regions of the ISM, since its 
global covering fraction is significant.

To estimate the galaxy escape velocity, we scale the 
\oii\ emission linewidths $\sigma$, which are an estimator of
the circular velocity $V_c$.  For a halo with 
outer radius $r_h$ and a flat rotation curve, 
the escape velocity at $R$ is $V_{esc}(R)^2 = 2V_c^2 {\rm ln}~(1+r_h/R)$
(Binney \& Tremaine 1987).
For reasonable halo mass distributions, 
$V_{esc} \simeq 3 V_c$, and is only weakly
dependent on the outer radius $r_h$.
The line-of-sight emission linewidths are related to the 
rotation velocity by $\sigma \sim (0.5~{\rm to}~0.6) V_c $, where
$\sigma$ incorporates an average over inclinations, as 
observed, but $V_c$ is the true value
(Rix \etal\ 1997; Kobulnicky \& Gebhardt 2000; Weiner \etal\ 2006).
There is scatter in this relation of $\sim 25\%$; in the mean 
it implies that 

\begin{equation}
V_{esc} \simeq (5~{\rm to}~6) \times \sigma(\oii).
\end{equation}
\noindent

\begin{figure}[t]
\begin{center}
\includegraphics[width=3.5truein]{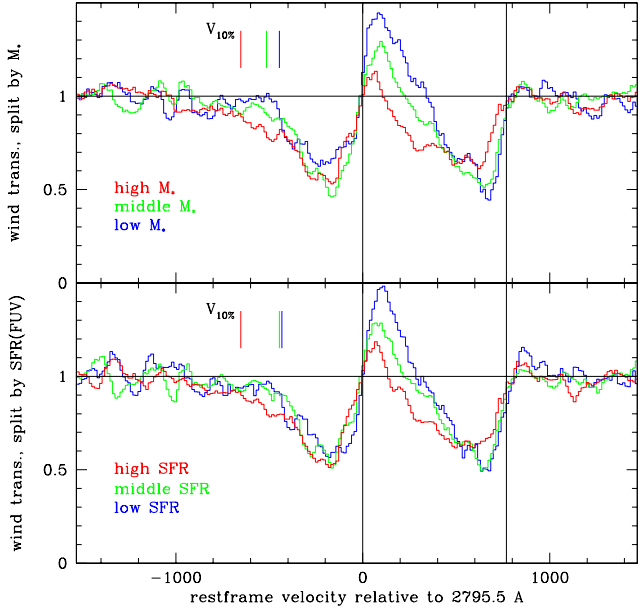}
\caption{
\mgii\ outflow absorption component $1-A_{flow}$
compared for galaxies in (upper panel) the three subsamples of stellar
mass; (lower panel) the three subsamples of $SFR_{FUV}$.  Velocity
is referred to \mgii\ 2795.5 \AA.
A 5 pixel boxcar smooth has been applied.  In each panel,
the red, green, blue lines are for high to low mass, or SFR.  The 
vertical lines at -400 to -650 \kms\ show the location of $V_{10\%}$,
where the outflow crosses 10\% absorption.
The high mass and high SFR galaxies show greater outflow absorption
at $-800 < V < -400$ \kms\ than the lower mass or SFR samples.
}
\label{fig-windcomp}
\end{center}
\end{figure}

\begin{figure}[t]
\begin{center}
\includegraphics[width=3.5truein]{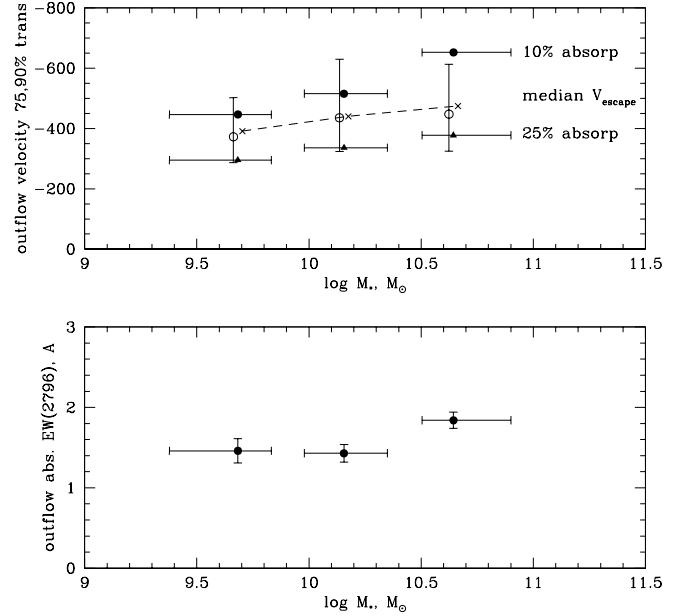}
\caption{
Upper panel: the outflow velocity at which 
wind \mgii\ absorption is 10\% and 25\% of continuum, for 
coadded spectra in three bins of stellar mass (filled points).  
The filled circles and horizontal error bars show the median 
and $\pm 34\%$ range of galaxies in the mass bin.  
The open circles and vertical errorbars are the median and $\pm 34\%$
range of escape velocity computed from the \oii\ linewidth of
galaxies in that bin, $V_{esc}=6\sigma(\oii)$.  The Xes are the 
average escape velocity from the \oii\ dispersion in the coadded
spectrum. Lower panel: the EW of outflow \mgii\ absorption in the 
coadded spectra in three bins of stellar mass.
}
\label{fig-velmass}
\end{center}
\end{figure}

\begin{figure}[t]
\begin{center}
\includegraphics[width=3.5truein]{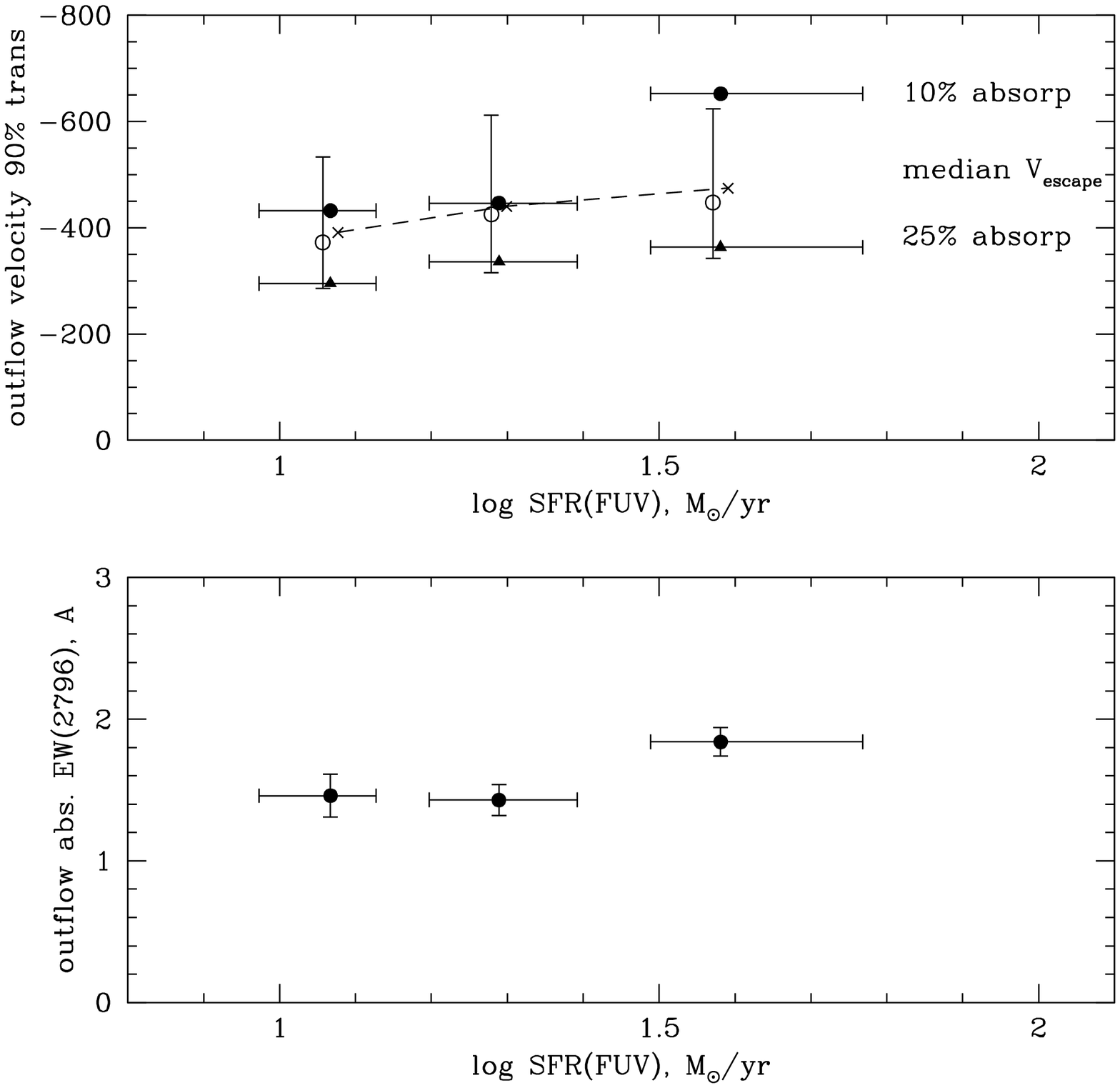}
\caption{
As Figure \ref{fig-velmass}, but for SFR(FUV).
Upper panel: outflow velocity at which wind \mgii\
absorption is 10\% and 25\% (filled circles) in three bins of SFR(FUV).
The open circles and vertical error bars are the 
median escape velocity $V_{esc}=6\sigma(\oii)$,
and the Xes are the average $V_{esc}$ from the coadded \oii\ linewidth.,
Lower panel: EW of outflow \mgii\ absorption in three bins
of SFR(FUV).
}
\label{fig-velsfr}
\end{center}
\end{figure}

To be conservative we have used the high factor of 6 in 
computing $V_{esc}$ for Figures \ref{fig-velmass} and \ref{fig-velsfr}.
These figures show $V_{esc}$ based on the composite \oii\ linewidth
and the median and $\pm34\%$ range of $V_{esc}$ based on the
individual linewidth measurements of galaxies in each
subsample.  Errors on the linewidths tend to
broaden the velocity distribution.
The threshold velocity
where \mgii\ absorption reaches 10\%, $V_{10\%}$, is moderately
higher than median $V_{esc}$
in the subsamples of low and intermediate mass and SFR.
However, in the high-mass or high-SFR subsample, $V_{10\%}$ is
substantially more than the median $V_{esc}$ estimated from the 
galaxy velocity widths.  This effect is caused by the tail of 
high-velocity absorption shown in the high-mass, high-SFR 
subsamples in Figure \ref{fig-windcomp}.

The high velocities in the tail of the outflow distribution
suggest that low-ionization gas is able to reach large radius or even 
escape.  This could give rise to \mgii\ halo absorption systems,
but more rigorous modeling of the EW distribution and covering
factor will be necessary to confirm that.  Prochter \etal\
(2006) have argued that the redshift evolution of
the number of strong \mgii\ absorbers links them to the
global SFR evolution and to
starburst events in low-mass galaxies.  
However, statistical detections of the associated light 
and the dust content of \mgii\ absorbers suggest a link to 
outflows from massive galaxies (Zibetti \etal\ 2007;
M\'enard and Chelouche 2008).

Bouch{\'e} \etal\ (2006)
advocated an outflow origin (rather than kinematics governed
by halo mass) for \mgii\ absorbers.  This was based on an 
anti-correlation between absorber equivalent width (velocity 
spread) and host halo mass, derived from clustering statistics.
However, we do find a correlation of outflow EW or velocity
spread with galaxy mass.  A more sophisticated treatment of the
clustering based on halo occupation suggests that the absorber-halo
mass relation is partly related to the shocking of cold gas in 
very massive halos (Tinker \& Chen 2008).
The ubiquity of \mgii\ outflows that
we find and the correlation of SFR and
outflow strength suggest that at least some fraction of \mgii\
absorbers arise in outflows from massive galaxies, but there is 
not yet a fully consistent link, and there could be
multiple causes for \mgii\ absorbers.

\subsection{Comparison to low-redshift winds and IR-luminous galaxies}
\label{sec-localcomp}

Given the galaxies' range of SFR from 10-100 \msunyr, 
and the median Petrosian radius of 5.2 kpc in the subset
with ACS imaging, essentially
all of the galaxies are above the criterion of SFR $>0.1$
\msunyrkpcsq\ where winds are typically found 
in the local universe (Heckman 2002).
This suggests that the criterion also applies to high
redshift galaxies, or conversely that it is no surprise
that winds are ubiquitous in this sample.  Because 
IR-luminous galaxies are more common at $z=1$ (Le Floc'h
\etal\ 2005) and the higher SFR at $z=1$ occurs across
the entirety of the starforming population (Noeske \etal\ 2007a),
in retrospect it might have been expected that winds should be
common at $z>1$. 

Martin (2005) established in low-z ULIRGs that the upper
envelope of the outflow velocity detected in Na I absorption 
depends weakly on SFR, $V_{wind} \sim SFR^{0.35}$.
As noted above, we find a similar power-law exponent,
$V_{wind} \sim SFR^{0.3}$, for the
$z=1.4$ star-forming galaxies, even though these are 
mostly well below ULIRG luminosities.
Rupke \etal\ (2005b) found a wide dispersion of outflow velocities
in galaxies with SFR $\gtrsim 100 \msunyr$, but a
correlation between outflow velocity and SFR or $V_c$ 
when both IR-luminous galaxies and dwarfs with winds are considered.
Their sample shows a weak correlation of Na I column with
SFR and galaxy mass.
Sato \etal\ (2008) use spectra from DEEP2 to study outflows
in Na I at $z\sim 0.6$, and find that the detection fraction
of winds is significantly higher in 24 \um\ detected galaxies
than in other similarly massive galaxies.  The Sato \etal\
sample does show some signs of outflow in galaxies without ongoing
star formation, a population we do not probe in this sample.
Our results agree
with the sense of these trends: higher SFR galaxies have deeper
and faster outflow absorption.

Comparing the $z \sim 1.4$ sample to the low-redshift
IR-luminous galaxies of Rupke \etal\ (2005b), the DEEP2 \mgii\
sample has SFR and \lir\ more like 
Rupke's IRGs ($L_{IR}>10^{11}\lsun$) than his ULIRGs; the median
DEEP2 galaxy is just at the $10^{11}\lsun$ threshold.
However, the $z \sim 1.4$ galaxies are somewhat more likely
to host winds and to have higher outflow velocities than the
low-redshift IRGs.  Our absorption depth for winds in DEEP2, 
$55\%$, corresponds to detection frequency times clumpiness
covering fraction in the Rupke IRGs,
which is lower at $C_\Omega C_f \simeq 0.2$.  
The velocity to which we detect winds is also higher.  Rupke \etal\
attempt to define a maximum velocity of robustly detected
absorption by $v_{max}$ containing $\sim 90\%$ of the absorption
and find the low-z IRGs have $v_{max} = 301^{+145}_{-98}$ \kms.
For comparison, we find that the velocity containing 50\% 
of absorption $V_{med}$ is often 250--300 \kms\ 
(Table \ref{table-absprops}), suggesting that $v_{max}$ is
much higher.  We find that at 10\% absorption depth,
$V_{10\%} = 400$ to $650$ \kms.

Rupke \etal\ (2005b) find almost
no absorbers at $>600$ \kms\ in star-formation dominated
galaxies, and we find significant column at that velocity.
It is quite possible that this comes from the difference 
between Na I and \mgii\ as tracers.  With its higher ionization
potential, \mgii\ should be easier to find in low column density
gas.  We also find that \mgii\ is close to fully saturated, with
a doublet ratio of 1.1 or even lower, while low-$z$ Na I studies find 
somewhat less saturated absorption at ratios 1.2-1.3 
(Heckman \etal\ 2000; Rupke \etal\ 2005b).  

The \mgii\ study of very luminous post-starburst galaxies at $z\sim 0.6$
by Tremonti \etal\ (2007)
found discrete absorbers at significantly higher outflow velocities 
than in our sample, often above 1000 \kms.  Those galaxies, being
post-starbursts are rather 
different from our sample and probably instances of AGN-driven rather 
than star formation driven winds.  AGN-driven winds are expected
to reach higher velocities than star formation-driven winds
observationally and theoretically 
(Veilleux \etal\ 2005; Scannapieco \& Oh 2004; Thacker \etal\ 2006).

Heckman \etal\ (2000) found relatively little dependence of
the outflow speed on host galaxy $V_c$, using a mix
of Na I and X-ray temperature measurements to obtain wind speed.
This result suggested 
that gas preferentially escapes from low-mass galaxies.
In contrast, in the $z\sim 1.4$ sample, we do find a
dependence of outflow speed on galaxy mass, possibly due to 
the more uniform use of \mgii\ as an outflow probe and the more 
homogeneous galaxy sample (versus the local samples which
tend to comprise IR-luminous galaxies and dwarfs with few
intermediate objects).

Rupke \etal\ (2005b) suggested 
that gas could escape from ULIRGs, and thus that massive
galaxies at high redshift influence the IGM; deep potentials
do not inhibit wind formation.  Martin (2005) 
also suggested that the dependence of velocity on SFR 
implies that wind feedback is important in high mass galaxies.
Our finding of outflow velocities at or greater than the escape
velocity, and a greater effect in the high-SFR subsample, reinforces
the point.  A large fraction of the global star formation
occurs in high-mass galaxies (e.g. Le Floc'h \etal\ 2005;
Noeske \etal\ 2007b); if outflows scale positively with
SFR, high-mass star forming galaxies have a 
significant impact on the enrichment and evolution of the IGM.

\subsection{Implications for wind models and dependence on mass}
\label{sec-windmodel}

These detections of the prevalence and properties of winds
in starforming galaxies can constrain both
the impact of winds on galaxies and the IGM, and the 
physical models of the wind mechanism.

Since the influential work of Larson (1974) it has been
recognized that winds driven by supernovae could arise 
during galaxy formation and limit its progress.  Larson 
suggested that these winds could cause more gas loss
in smaller galaxies and explain the mass-metallicity
relation.  Subsequent treatments have emphasized the
importance of winds for solving problems with low-mass
galaxies in cold dark matter models of structure formation,
such as reducing the number, luminosity, and metallicity
of faint galaxies (e.g. Dekel \& Silk 1986; Dekel \& Woo 2003).

Hydrodynamical studies have shown that while it is relatively
difficult for supernovae-driven outflows to clear low-mass
galaxies entirely of gas, they can be fairly efficient at ejecting
metals (De Young \& Heckman 1994; Mac Low \& Ferrara 1999).
Star-formation driven winds are a necessary ingredient,
under the name of ``feedback,'' for models of galaxy formation
to reproduce observed scaling relations (e.g. Somerville \& 
Primack 1999; Ceverino \& Klypin 2007).  Matching the
observed radius, luminosity and velocity relations of
disk galaxies requires that a large fraction of baryons
either never make it into the disk or are blown out
(Dutton \etal\ 2007).

Models of winds have tended to concentrate on the effect
of winds on low-mass galaxies, where the question of complete 
gas blowout is more pressing, and where outflows are effective
at leaving fossil evidence in the mass-metallicity relation
(Garnett 2002; Tremonti \etal\ 2004; Dalcanton 2007).
Star-formation driven outflows do not leave a strong trace in
the metallicity of high-mass, low gas fraction galaxies, and so
the mass-metallicity relation is not evidence against winds
in high-mass galaxies (Dalcanton 2007).  The baryonic 
Tully-Fisher relation does not show a break at low masses
(Geha \etal\ 2006), as might be expected in models where
winds are effective below a critical mass or circular
velocity (e.g. Dekel \& Silk 1986).
Some models have predicted that 
the effect of winds is small in high-mass galaxies because
the wind is blocked by gas within the disk
(e.g. Dubois \& Teyssier 2008; Dalla Vecchia \& Schaye 2008).

But in fact, high-velocity winds in high-mass,
SFR-dominated galaxies are demonstrated to exist
in Lyman break galaxies (Shapley \etal\ 2003), in
LIRG/ULIRGs (Rupke \etal\ 2005b; Martin 2005), and 
in blue-cloud starforming galaxies (this work).  These could 
have significant effects on gas content, absorption-line systems, 
and the enrichment of the IGM and ICM, and so proper treatment
of winds in high-mass galaxies is necessary to model these
problems.  Although winds probably have a stronger influence
on the structure of low-mass galaxies than on high-mass galaxies,
this does not tell us whether low- or high-mass galaxies 
are more important for IGM enrichment.

The behavior of winds in low and high mass galaxies is
related to another theoretical problem, the appropriate
physical model for the driving of the outflow.  
The cool gas in winds may be driven either by the kinetic energy of 
supernova ejecta, by entrainment in the hot wind
(e.g. Chevalier \& Clegg 1985; Heckman \etal\ 1990;
Strickland \& Stevens 2000)
or by the momentum, through radiation pressure
(Murray \etal\ 2005) or cosmic ray pressure (Socrates \etal\ 2006).
These predict different scalings for the behavior of wind
velocity with galaxy mass and SFR.

The observed X-ray temperature of winds varies little with galaxy
mass, which would suggest relatively little dependence of
cool gas velocity on galaxy mass
(Martin 1999; Heckman \etal\ 2000).  As pointed out by 
Martin (2005), the increase of wind velocity with SFR, 
$V \sim SFR^{0.35}$, which she finds in ULIRGs and we find
across a broad range of starforming galaxies, is surprising
if winds are energy-driven (though see Fujita \etal\ 2008).  
Momentum-driven winds appear to
fit this scaling better (Murray \etal\ 2005),
although a detailed comparison is beyond the scope
of this paper.  

We discuss one possible implication,
the shape of the outflow absorption
profile.  In the momentum-driven wind model of Murray \etal\ (2005), 
cool gas is accelerated by radiation pressure, predicting
$V^2(r) = 4\sigma_h^2 (L/L_M-1) {\rm ln}~ (r/r_0)$,
where $\sigma_h$ is the halo velocity dispersion and 
$L_M$ is a critical luminosity to drive a wind.
The radial dependence of the wind-driven cool gas
density is not easy to predict, but it should 
decline with radius, e.g. $\rho(r) \sim r^{-2}$ to $r^{-3}$.

\begin{figure}[t]
\begin{center}
\includegraphics[width=3.5truein]{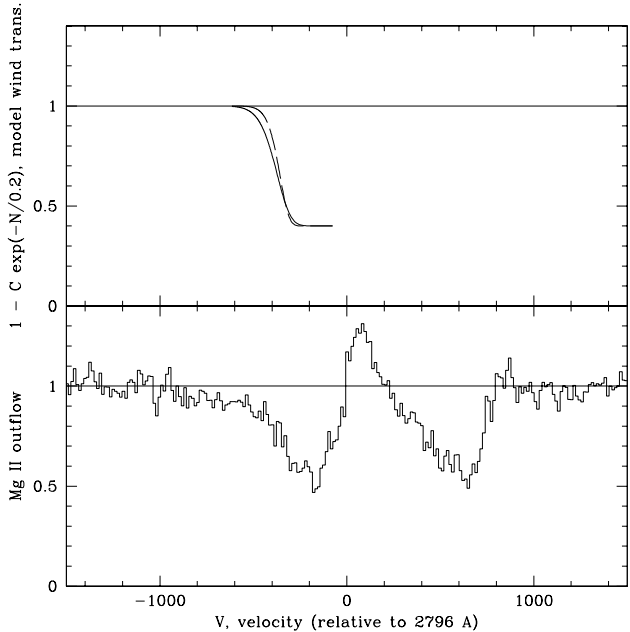}
\caption{
Upper panel: Absorption profile predicted by a
model in which the gas is accelerated, with $v(r)$ increasing
and $\rho(r)$ decreasing (e.g. Murray \etal\ 2005). 
The optical depth and covering factor
have arbitrary normalization; the covering factor is 60\%.
The solid line is for $\rho \propto r^{-2}$ and the dashed
line for $\rho \propto r^{-3}$.  This model can produce
an asymmetric profile.
Lower panel: Observed \mgii\ outflow 
component profile, velocity relative to 2795.5 \AA, in the
no excess emission sample, with an asymmetric tail to
high outflow velocity.
}
\label{fig-windmodel}
\end{center}
\end{figure}

Figure \ref{fig-windmodel} shows a toy model of the 
absorption that such a wind might produce (upper panel)
for $\sigma_h=150$ \kms\ and a 60\% covering factor,
to be compared with the \mgii\ outflow (lower panel).
We have simply convolved the equations for $V(r)$ and $\rho(r)$,
assuming that the total covering factor is 60\% and the 
absorption strength declines as $\rho(r)$ (lower covering factor 
and/or optical depth as the gas spreads out to large radius).
This model is by no means unique; the essential
point is that a model in which the cool gas is accelerated
as radius increases, while its density declines with $r$,
can naturally produce an asymmetric line profile with a tail of
absorption to high velocities.  This shape is
similar to the composite spectrum and to the individual
absorption lines we detect (Figure \ref{fig-samplespec}).
The model spectrum does not have gas
at velocities as large as observed, which could be due
to limitations of the model and our assumptions about $\sigma_h$
and $L_M$.  This is not a unique model prediction; for
example, low column density clouds might be accelerated 
to higher velocities (see Rupke \etal\ 2005b), possibly
producing a similar profile. 
Detailed comparisons of wind models
and the spectra should be a constraint on wind physics.

Wind velocities that increase with mass and SFR undermine
some of the assumptions in models of winds
that emphasize blowout in low mass galaxies (e.g. Dekel \& Silk 1986).
Incorporating wind velocities that increase with mass
($V_{wind} \sim V_{esc}$, rather than constant $V_{wind}$)
into cosmological simulations does a better job of
reproducing the IGM enrichment and mass-metallicity
relation (Oppenheimer \& Dav\'e 2006; Finlator \& Dav\'e 2008).
Regardless of the energy vs. momentum-driven controversy, 
we find that indeed $V_{wind} \sim V_{esc}$ (Figures \ref{fig-velmass}
and \ref{fig-velsfr}), reinforcing that this is a better
scaling to apply.


\subsection{Descendants of the DEEP2 sample}
\label{sec-descend}

Winds have previously been detected in low-redshift dwarf
starbursts and IR-luminous galaxies (e.g. Heckman \etal\ 2000;
Martin 2005; Rupke \etal\ 2005b), in $z\sim 0.6$ luminous
post-starbursts (Tremonti \etal\ 2007), and in $z\sim 3$
Lyman break galaxies (Pettini \etal\ 2000; Shapley \etal\ 2003).
A common feature of all but the dwarfs is that they have a 
high fraction of mergers or progenitors of ellipticals; in the case
of the $z\sim 3$ galaxies this is based on their clustering
(Adelberger \etal\ 2005).  This leaves open the question
of whether winds were mostly associated with elliptical 
formation or were also part of the past of today's
luminous non-starburst spirals.  Were outflows
driven from massive galaxies only in bursty events associated
with mergers or quenching of star formation, or were they 
also driven during the relatively steady high-SFR mode 
that governs blue galaxies at $z \sim 1$ (Noeske \etal\ 2007a)?

The DEEP2 \mgii\ sample shows that winds are common in the
blue cloud of star-forming galaxies at $z\sim 1.4$.  
Although we are only probing the brightest 1-1.5 mag 
of galaxies at a given $U-B$ color, the sample spans
a wide range of stellar mass and color, showing that
winds occur even in low-mass galaxies with $M_*<10^{10}\ \msun$.
These are well below the transition mass for quenching
of star formation (Bundy \etal\ 2006) and are unlikely
to evolve quickly onto the red sequence.

Because the DEEP2 \mgii\ sample is the high-redshift tip of
a classical magnitude-selected survey, we can trace the
blue population down in redshift.  The evolution of the
blue cloud in DEEP2 suggests that most of the galaxies 
stay blue and fade as their SFR declines gradually (Blanton 2006;
Willmer \etal\ 2006; Faber \etal\ 2007; Noeske \etal\ 2007a).
Some of the brightest blue galaxies may turn red; 
especially  in denser environments.
At $z>1$, some of the brightest blue DEEP2 galaxies are in
dense environments (Cooper \etal\ 2007), and from 
$z=1.3$ to $z=1$, the red fraction in DEEP2 groups increases
(Gerke \etal\ 2007).  It is likely that some of the $z \sim 1.4$
DEEP2 blue galaxies will turn into ellipticals
in groups.  However, much of the sample is in the field;
blue galaxies in DEEP2 have high number density and
much lower clustering
than local early-type galaxies (Coil \etal\ 2008).
The combination of number density and 
clustering shows that most of them will not evolve into
local early-types, similar to the argument made by Conroy \etal\ (2008)
for the descendants of $z\sim 2$ galaxies.
Note also that of the galaxies with HST/ACS imaging, the
merger fraction was very low, 3/118, while 50/118 were
classified as face-on or inclined disky objects (Section
\ref{sec-typedepend}).

An analysis of the clustering and number density of 
$z \sim 2$ star-forming galaxies indicates that they
will evolve mostly into $L_*$ galaxies today, including
many massive spirals and some ellipticals (Conroy \etal\ 2008).
The $z \sim 2$ galaxies are roughly 
intermediate in properties and time between the $z\sim 3$
Lyman break galaxies and the $z\sim 1.4$ DEEP2 galaxies,
reinforcing that the DEEP2 \mgii\ galaxies are most
likely to turn into massive spirals like the Milky Way.
The lookback time to $z=1.4$ is 9.0 Gyr, approximately the
epoch at which the Milky Way thin disk begins to form 
(Edvardsson \etal\ 1993), which could 
plausibly be associated with a very high SFR.  

We conclude
that the prevalence of winds in the DEEP2 $z \sim 1.4$
sample shows that many $L_*$ galaxies, including spirals,
drove a wind during their early, high-SFR, formative phases.
Most blue galaxies continue to form stars down
to the present day, which implies that a strong SFR-driven wind 
does not prevent subsequent star formation.

Winds, possibly from AGN, are a popular mechanism for truncating or
quenching further gas accretion and star formation during merger
formation of spheroidal galaxies (e.g. Sanders 1988; Cox \etal\ 2006;
Tremonti \etal\ 2007).  The physics of how the energy
source couples to the ISM to clear it from the galaxy are not
well understood.  Galaxy-scale simulations cannot resolve the AGN,
and popular models incorporating AGN feedback 
(e.g. Hopkins \etal\ 2006) couple the AGN energy directly to the
ISM, which could be unrealistically efficient.

It may be that AGN energy input, orbital energy from the merger, or an
extremely high SFR (effectively Eddington limited, as in
Murray \etal\ 2005 or Socrates \etal\ 2006) 
can clear the ISM from the progenitors of ellipticals, but these
have yet to be proven.  
SFRs like those observed in the DEEP2 $z \sim 1.4$ sample, 10-50 \msunyr,
do not appear to remove enough gas to deter subsequent
star formation.
There is evidence that radiation from QSOs
heats and ionizes the IGM out to large distances, as seen in the
proximity effect (shown for metal lines by Wild \etal\ 2008), 
but also that the effect is anisotropic, weaker in the transverse 
direction (Hennawi \& Prochaska 2007).  The physics of QSOs
ionizing the ISM or IGM absorbers are understood, but it is 
more difficult to couple radiation to the gas to transfer momentum 
and cause blowout, which is a current weakness of the AGN
quenching scenario.

\section{Conclusions}
\label{sec-conclusions}

We have used 1406 spectra of blue star-forming
galaxies at $z \sim 1.4$
in the DEEP2 redshift survey to detect blueshifted \mgii\ 
and \mgi\ absorption in the coadded spectrum, and in a
subset of the individual spectra that have high enough
S/N.  The blueshifted \mgii\ absorption reaches a depth
of 55\%, indicating that large-scale
wind-driven outflows in cool, low-ionization gas
exist in at least half of the galaxies.  
The convolution of frequency and covering fraction of outflows in
low-redshift IR-luminous galaxies is $\sim 0.2$ (Rupke \etal\ 2005b).
The 55\% absorption in our sample suggests that in fact all of the
$z \sim 1.4$ star-forming
galaxies are driving outflows, and that the \mgii\ covering fraction
is higher than seen in Na I in the local IR-luminous galaxies.

The outflows are driven by star formation;
about 50 of the galaxies show narrow \mgii\ emission indicating
AGN activity, but these have a similar outflow signature to the rest.
The outflows show an asymmetric absorption profile with a
tail to high velocities; the median of the absorption is
$\sim -250$ \kms, with absorption at 10\% depth out to 
$\sim -500$ \kms.  In the high stellar mass subsample, absorption is
visible out to $\sim -1000$ \kms.

The \mgii\ absorption is quite saturated and the overall
covering fraction is large, given the 55\% absorption trough.
Applying the doublet ratio method to the \mgii\ doublet,
we find that the optical depth is $\sim 10$ and the column
density of the absorbers is $N_H \sim 10^{20}$, possibly
even higher given the high optical depths.  This implies
that a typical luminous star-forming galaxy at $z=1.4$
is blowing out and covered by a sub-damped Lyman-$\alpha$
absorber.  

The sample galaxies have SFRs of $10-100$ \msunyr, derived
from UV fluxes and normalized to infrared luminosities; the median
galaxy has $\lir = 1.3 \times 10^{11}~\lsun$.
We find mass outflow rates of order 20 \msunyr\ using the
measured cool gas outflow speed and column density, and 
a radius of at least 5 kpc, about the optical size of the
galaxies.  The outflow rate is of the same order as the
galaxies' star formation rate, in accord with local starburst wind
galaxies and models of high-redshift chemical evolution.
Maintaining both high SFR and high outflow suggests that galaxies
process gas relatively quickly and the rates are governed
by infall (e.g. Heckman \etal\ 2000; Erb 2008).  Substantial outflows
can explain why the baryon fraction in galaxy disks is lower
than the global value.
The actual radius the outflows can reach is
not known; it is possible that these outflows will give
rise to \mgii\ absorption line systems at 20-50 kpc.

We divide the galaxies into subsamples by stellar mass, color, and
star formation rate.  We find strong blueshifted \mgii\
absorption in all subsamples, across factors of $\sim 30$
in stellar mass and $\sim 10$ in SFR.  For 118 of the galaxies,
we have HST/ACS imaging and divide the galaxies into
types including disks, irregulars and mergers.  Again
we find outflow absorption in all types.  Nearly half
of the galaxies are spirals and only 3/118 are obvious
mergers.  This is distinct from wind samples studied in absorption
at lower redshifts, which tend to be either IR-luminous
starbursts (Heckman \etal\ 2000), IR-luminous galaxies
with a large merger fraction (Rupke \etal\ 2005b; Martin 2005),
or post-starbursts (Tremonti \etal\ 2007).
The blue galaxy population has a high SFR across all
luminous galaxies at $z>1$ (Noeske \etal\ 2007a), rather than
high SFR confined to a small fraction of the objects, and
this places many of the blue galaxies at $z\sim 1.4$ above the
$SFR > 0.1 $ \msunyrkpcsq\ criterion for driving a wind that
appears to hold in the local universe (Heckman 2002).

The \mgii\ outflow absorption
is stronger and reaches to larger velocities for
high-stellar-mass, high-SFR galaxies.  The outflow velocity 
scales with SFR as
$V_{wind} \sim SFR^{0.3}$, similar to the scaling found
by Martin (2005) in low-redshift ULIRGs.
The outflow velocity scales roughly as the galaxy escape 
velocities, which we derive from \oii\ emission line velocity
widths.  Although theoretical treatments of galactic
winds have tended to emphasize their effect on low-mass 
galaxies, the velocity scaling we find suggests that 
gas may be able to escape from high-mass galaxies. 
A scaling relation of this type should be applied in theoretical
models of galaxy formation and IGM evolution 
(e.g. Finlator \& Dav\'e 2008).
This velocity scaling may support momentum-driven wind
models over energy-driven winds;
further study of the wind models and the detailed 
absorption line profiles should yield insights into
the wind physics.

Most of these $z \sim 1.4$ DEEP2 blue star-forming galaxies will
evolve into massive $L_*$ galaxies at $z=0$, including
many star-forming spirals,
although some fraction will turn into ellipticals.  The prevalence
of \mgii\ outflows in this DEEP2 sample thus indicates that
winds are not confined to mergers and elliptical progenitors;
the progenitors of Milky Way-type galaxies drove winds in
the past.  Traces of these past winds in high-mass galaxies
are likely not visible in the mass-metallicity relation
(Dalcanton 2007), but
they will have influenced the chemical evolution of the 
galaxies and the IGM.

Most of the descendants of the blue star-forming galaxies
continue to form stars, which implies that the winds in these
galaxies are not strong enough to clear the ISM and quench
star formation.  Powerful winds during galaxy mergers are
a popular mechanism for truncating gas accretion and star
formation in red galaxies (e.g. Sanders 1988; Cox \etal\ 2006).
These invoke energy 
input from AGN, orbital energy of the merger, or extremely high
star formation rates.  The details of how the energy source
couples to the ISM to clear it are not well understood.
We show that strong star-formation driven winds do not deter
subsequent star formation in the blue galaxy population.
Although star-formation driven winds are not expected to
clear their hosts' ISM based on their energetics, 
AGN-driven winds must be consistently stronger and very
well coupled to the ISM if they are to explain quenching.
The physics of the coupling are a difficult problem that
must be understood to make wind quenching scenarios viable.

\medskip

\acknowledgments

We are grateful to Christy Tremonti for inspiring this project
and much advice, and to Kristian Finlator,
Romeel Dav\'e, David Rupke, Sylvain Veilleux and Norman Murray 
for helpful discussions.
We acknowledge the cultural role of the summit of Mauna Kea within the
indigenous Hawaiian community and are grateful to have been able to
observe from this mountain.  
BJW has been supported by NASA/Spitzer contract 1255094.
ALC is supported by NASA through Hubble Fellowship grant
HF-01182.01-A awarded by STScI, which is operated by AURA for NASA
under contract NAS 5-26555.
DEEP2 has been supported by NSF grants AST05-07428 and AST05-07483.
WIRC observations were obtained at the Hale Telescope, Palomar 
Observatory as part of a continuing collaboration between Caltech,
NASA/JPL, and Cornell University.  This work is based in part on
observations made with the Spitzer Space Telescope 
operated by and funded through the Jet Propulsion Laboratory, 
California Institute of Technology under a contract with NASA.  
Support for this work was provided by NASA through contract
1255094 issued by JPL/Caltech.

{\it Facilities:} \facility{Keck}, \facility{Spitzer (MIPS)}, \facility{Hale}.

\appendix

\section{Stellar masses in DEEP2 from the color--$M/L$ relation}
\label{app-mstellar}

It has been shown for local galaxies that reasonable 
star formation histories produce a relation between 
color and stellar $M/L$ (Bell \& de Jong 2001).  We
would like to use such a relation to infer stellar mass
$M_*$ for galaxies that do not have $K$-band data nor
stellar masses computed by Bundy \etal\ (2006).  We should
not directly apply the local relations (Bell \& de Jong 2001; 
Bell \etal\ 2003) since higher-redshift galaxies have
different star formation histories, so we empirically
correct the color-$M/L$ relations using galaxies that
do have $K$-band stellar masses.

We first compute the relation for $M_*/L_B$ as a
function of $B-V$ for $z=0$ galaxies (Bell \etal\ 2003):

\begin{equation}
{\rm log}~M/L_B(z=0) = -0.942 + 1.737 ~ (B-V_{Vega}),
\end{equation}

\noindent
for a ``diet Salpeter'' IMF\footnote{For many typical galaxy SF
histories, the ``diet Salpeter'' IMF implies about 0.15 dex lower 
mass than a Salpeter IMF, a Kroupa IMF is slightly
lower than diet Salpeter, and a Chabrier IMF is 0.2-0.25 dex 
lower than Salpeter.  In practice, for our sample, the main 
effect of different IMFs is to shift the zeropoints of mass or SFR
due to a change in the ratio of high to low mass stars. 
SFR calibrations are often referenced to the Salpeter IMF 
(e.g. Kennicutt 1998) or sometimes Kroupa (Bell \etal\ 2005), 
while stellar mass estimates often use somewhat lighter IMFs
(Bell \& de Jong 2001; Bundy \etal\ 2006).  It is vitally 
important when comparing SFR estimates to compensate for the
assumed IMF.}
(Bell \& de Jong 2001).  
$M_B$ and $B-V$ are obtained by
our K-correction methods.  

This color-$M/L$ relation could easily
change with redshift.  We calibrate the stellar masses
by a least-squares fit of a correction term $C_K$ between the
color-$M/L$ masses and the $K$-derived masses 
of Bundy \etal\ (2006).  We fit to all 11924 galaxies with $K$
and masses in $0<z<1.5$, not just the \mgii\ sample, and find:

\begin{equation}
M_* = L_{B,Vega} \times  M/L_B(z=0) \times C_K(U-B,z),
\end{equation}

\begin{equation}
{\rm log}~C_K(U-B,z) = -0.0244 -0.398\ z + 0.105\ (U-B_{Vega}).
\end{equation}

\noindent
Here $U-B_{Vega} = U-B_{AB}-0.85$ and $B-V_{Vega} = B-V_{AB}+0.11$.
The scatter about the fit is 0.25 dex in stellar mass.  This
formula with $C_K$ applies to the entire DEEP2 sample over the 
range $0<z<1.5$.  There
are systematic residual patterns with color and redshift
of 0.1 dex (0-to-peak) in the mean, but this is less than the
0.3 dex rms scatter of individual galaxies about the relation.  
Introducing second-order terms
in  $U-B$ and $z$ does not improve the residuals significantly.

The corrected stellar masses presented here are
on the scale of Bundy \etal\ (2006), who used a Chabrier IMF.
For the 1406-galaxy \mgii\ sample, 409 have $K$ magnitudes
and masses from Bundy \etal\ (2006).  Figure \ref{fig-kmasscomp}
compares the mass estimates from color-$M/L$ to those from $K$-band,
and demonstrates that there is a good correlation, although
with significant scatter: the color-$M/L$ masses are 0.09 dex 
lower with an RMS dispersion of 0.30 dex.

\end{document}